\documentclass[12pt,preprint]{aastex}

\usepackage{color}

\def\gsim{\;\lower4pt\hbox{${\buildrel\displaystyle >\over\sim}$}\;}
\def\lsim{\;\lower4pt\hbox{${\buildrel\displaystyle <\over\sim}$}\;}
\def\grls{\;\lower4pt\hbox{${\buildrel\displaystyle >\over <}$}\;}

\shortauthors{WANG ET AL.} \shorttitle{EVE Thermodynamic Spectrum}

\begin{document}

\title{Thermodynamic Spectrum of Solar Flares Based on SDO/EVE Observations: Techniques and First Results}

\author{Yuming Wang$^{1,2}$, Zhenjun Zhou$^1$, Jie Zhang$^3$, Kai Liu$^{1}$, Rui Liu$^{1,4}$, Chenglong Shen$^{1,2}$,
and Phillip C. Chamberlin$^{5}$}

\altaffiltext{1}{CAS Key Laboratory of Geospace Environment,
Department of Geophysics and Planetary Sciences, University of Science and
Technology of China, Hefei, Anhui 230026, China (Email: ymwang@ustc.edu.cn)}
\altaffiltext{2}{Synergetic Innovation Center of Quantum Information \& Quantum Physics, University of Science and Technology of China, Hefei, Anhui 230026, China}
\altaffiltext{3}{School of Physics, Astronomy and Computational Sciences,
George Mason University, 4400 University Drive, MSN 6A2, Fairfax, VA 22030, USA}
\altaffiltext{4}{Collaborative Innovation Center of Astronautical Science and Technology, Hefei 230026, China}
\altaffiltext{5}{Solar Physics Laboratory, Heliophysics Division, NASA Goddard Space Flight Center, Greenbelt, MD 20771, USA}

\begin{abstract}
SDO/EVE provides rich information of the thermodynamic processes of
solar activities, particularly of solar flares. Here, we develop a
method to construct thermodynamic spectrum (TDS) charts based on the
EVE spectral lines. This tool could be potentially useful to the EUV
astronomy to learn the eruptive activities on the distant astronomical
objects. Through several cases, we illustrate what we can learn
from the TDS charts. Furthermore, we apply the TDS method to 74 flares equal
to or greater than M5.0-class, and reach the following statistical results.
First, EUV peaks are always behind the soft X-ray (SXR) peaks and stronger flares
tend to have a faster cooling rate. There is a power law correlation
between the peak delay times and the cooling rates, suggesting a coherent
cooling process of flares from SXR to EUV emissions. Second, there are
two distinct temperature drift patterns, called Type I and Type II. For Type I
flares, the enhanced emission drifts from high to low temperature like
a quadrilateral, whereas for Type II flares, the drift pattern looks like a
triangle. Statistical analysis suggests that Type II flares are more
impulsive than Type I flares. Third, for late-phase flares, the peak
intensity ratio of the late phase to the main phase
is roughly correlated with the flare class, and
the flares with a strong late phase are all confined. We believe
that the re-deposition of the energy carried by a flux rope, that
unsuccessfully erupts out, into thermal emissions is responsible for
the strong late phase found in a confined flare. Besides, with some cases
we illustrate the signatures of the flare thermodynamic process in the 
chromosphere and transition region in TDS charts. These results provide
new clues to advance our understanding of the thermodynamic processes
of solar flares and associated solar eruptions, e.g., coronal mass
ejections.
\end{abstract}

\section{Introduction}

As one of the most catastrophic events on the sun,
solar flares directly impact the environment of interplanetary space and the Earth's
atmosphere. During a flare process, free magnetic energy is converted into
electromagnetic radiation, energetic particles, heated plasma, waves, etc~\citep[e.g.,][]{Hudson_2011}.
The radiation occupies the majority of the flare
energy~\citep{Emslie_etal_2012}. Extreme-Ultraviolet (EUV) wavelengths, from about 10 to
120 nm, are a main window to observe solar activities~\citep{Frohlich_Lean_2004}.
Although it occupies a small fraction of solar total irradiance, the majority of
the Sun's variability appears in the EUV output~\citep[e.g.,][]{Woods_etal_2006, Moore_etal_2014}.

The history of solar EUV observations could be traced back to sounding rocket and satellite
experiments~\citep{Friedman_1963}. After that, many space missions, e.g., SOHO, TRACE, RHESSI, Hinode,
STEREO and SDO, made significantly observational achievements using their imaging
spectrographs with high spatial resolution and EUV broadbands. Except the EUV observations for the
Sun, there are also lots of EUV observations for stellar sources, such as those by EUVE satellite,
which acquired data in the wavelength from about 7 to 76 nm~\citep{Craig_etal_1997}. 

One of the most recent space-borne instrument for the solar EUV observations is the 
EUV Variability Experiment (EVE, \citealt{Woods_etal_2012}) on board the Solar Dynamics Observatory
(SDO, \citealt{Pesnell_etal_2012}), which has an unprecedented high cadence
of 10s, subtle spectral resolution of 0.1nm and breakthrough wavelength range
from 5 to 105 nm. Due to its excellent performance, some new features of the solar irradiance
are revealed~\citep[e.g.,][]{Woods_etal_2011, Chamberlin_etal_2012, Milligan_etal_2012, Milligan_etal_2012a,
Warren_etal_2013, LiuK_etal_2013, LiuK_etal_2015, Ryan_etal_2013}, particularly on the aspect of solar flares.
For example, a new phase of flares, called late phase,
was found after the flare's main phase at warm coronal lines~\citep{Woods_etal_2011}.
The joint analysis with the imaging data from SDO Atmospheric Imaging Assembly (AIA, \citealt{Lemen_etal_2012})
suggested that the further enhancement of emissions at
warm coronal lines during the late phase is associated with the heating of separately coronal
loops immediately near the main flare loops~\citep[e.g.,][]{Woods_etal_2011, LiuK_etal_2013}. \citet{Chamberlin_etal_2012}
and \citet{Ryan_etal_2013} studied the thermal evolution and radiative output of flares.
\citet{Warren_etal_2013} argued that the isothermal postulate seems unreasonable for the
thermal structure of a flare through comparing EVE
spectra with calculations based on parameters derived from the GOES soft X-ray (SXR) fluxes.

Each emission line in EUV is produced by a particular element in a particular ionization level that
corresponds to a formation temperature. Thus, EVE data with its high resolution in both wavelength and time
provide us a unique opportunity to study the thermal dynamics of solar activities, though
it does not have spatial resolution. So far, most studies of EVE data investigated the temporal profile of each
individual spectral line, which is not efficient and may miss some interesting features.
In this paper, we develop a method to construct the EVE thermodynamic
spectrum (TDS) chart, a 2-dimensional (2D) image of emission line intensity or other relevant quantities against
the temperatures (along Y-axis) and the time (along X-axis).
This is similar to the dynamic spectrum of radio emission, which is a 2D image of radio emission
intensity against the frequency and time. As will be seen below,
the charts could provide a global view of the thermal process of solar activities,
particularly of solar flares, reflected in the EUV wavelengths. The description of the 
data and method are given in the next section.

\section{Data and method}\label{sec_method}

\subsection{Selection of emission lines}\label{sec_line}
EVE instrument has three subsystems, among which MEGS (Multiple EUV Grating Spectrograph)
measures the spectral irradiance from 5 to 105 nm with 0.1 nm spectral resolution and with
10-second cadence~\citep{Woods_etal_2012}. MEGS has four channels: MEGS-A, MEGS-B, MEGS-SAM and MEGS-P.
Our study is based on the data from MEGS-A and MEGS-B, which were designed for the wavelength ranges of
5--37 nm and 35--105 nm, respectively. MEGS provide several level 2 data products, including the `line' (EVL)
product and the `spectra' (EVS) product. The EVL product consists of 30 emission lines;
half of them are extracted from MEGS-A EVS data and the other half from MEGS-B EVS data
(refer to the readme file at the official website
of EVE, \url{http://lasp.colorado.edu/home/eve/science/instrument/}).
MEGS-A has the full coverage in time (except the eclipse time for the SDO), whereas MEGS-B
does not operate full time and in the most time it only opened for about 5 minutes every hour.
Although the MEGS-A channel was lost on 26 May 2014, almost 5 years of data have been acquired and
thousands of flares have been observed, to which the TDS analysis can be applied.
We first use the data from only MEGS-A to show how to construct the TDS charts in this section,
and present the case and statistical studies on solar flares based on the TDS in Sec.~\ref{sec_cases} -- \ref{sec_results}. 
Then we show the extended-TDS charts constructed by combining both MEGS-A and MEGS-B data in Sec.~\ref{sec_extended},
as the sporadic MEGS-B data also recorded hundreds of flares.

\begin{figure}[tbh]
  \centering
  \includegraphics[width=0.32\hsize]{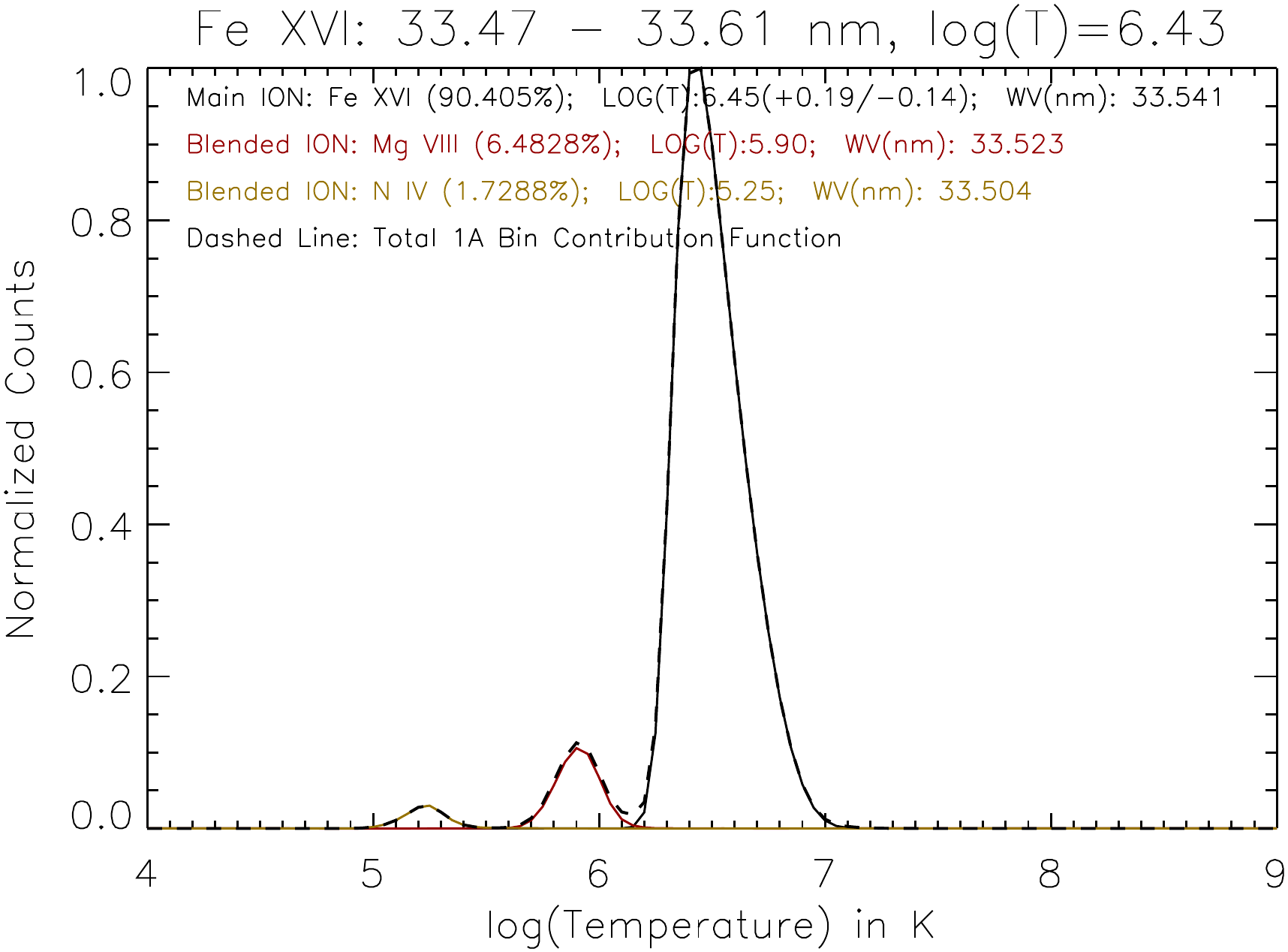}
  \includegraphics[width=0.32\hsize]{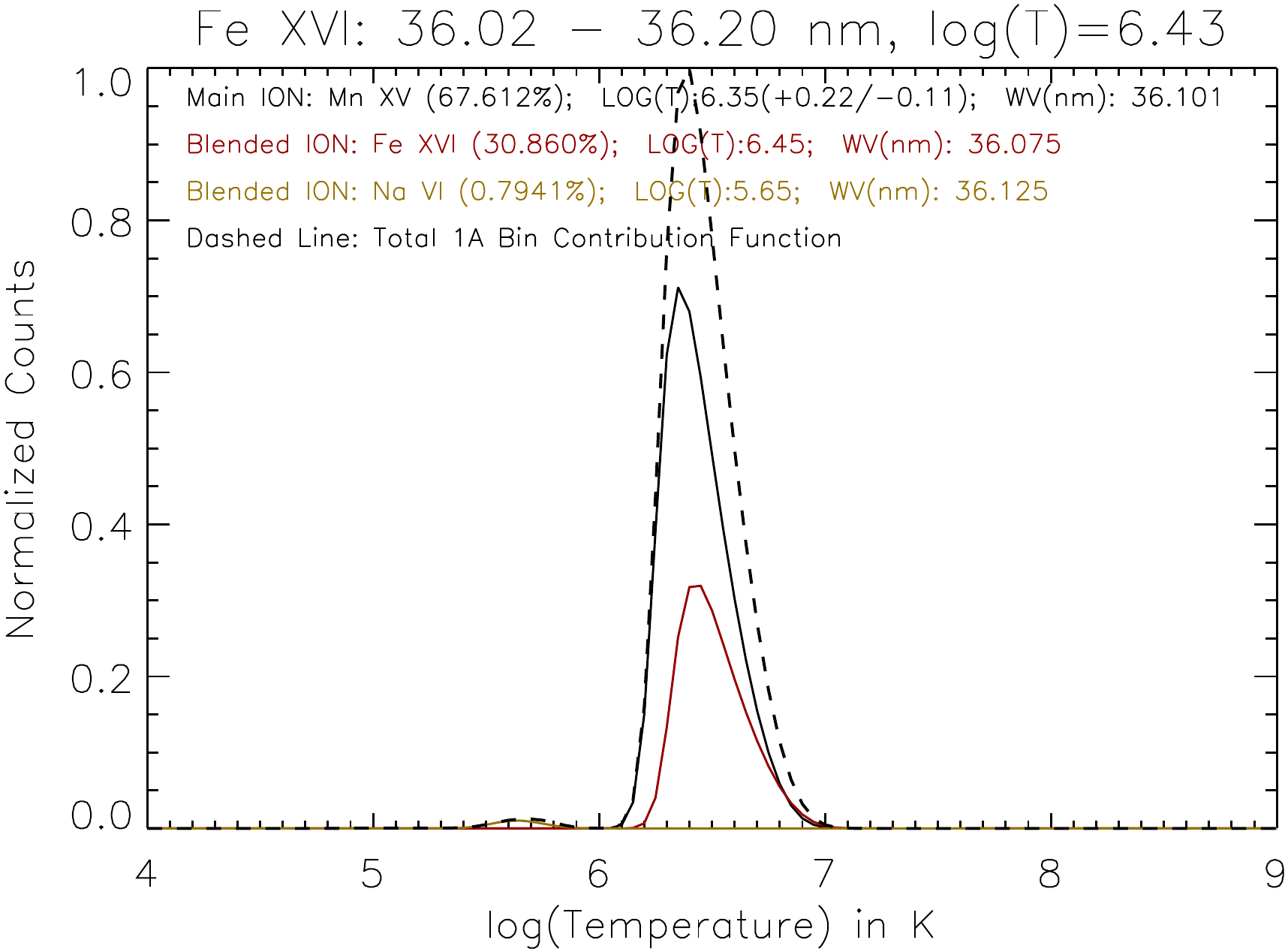}
  \includegraphics[width=0.32\hsize]{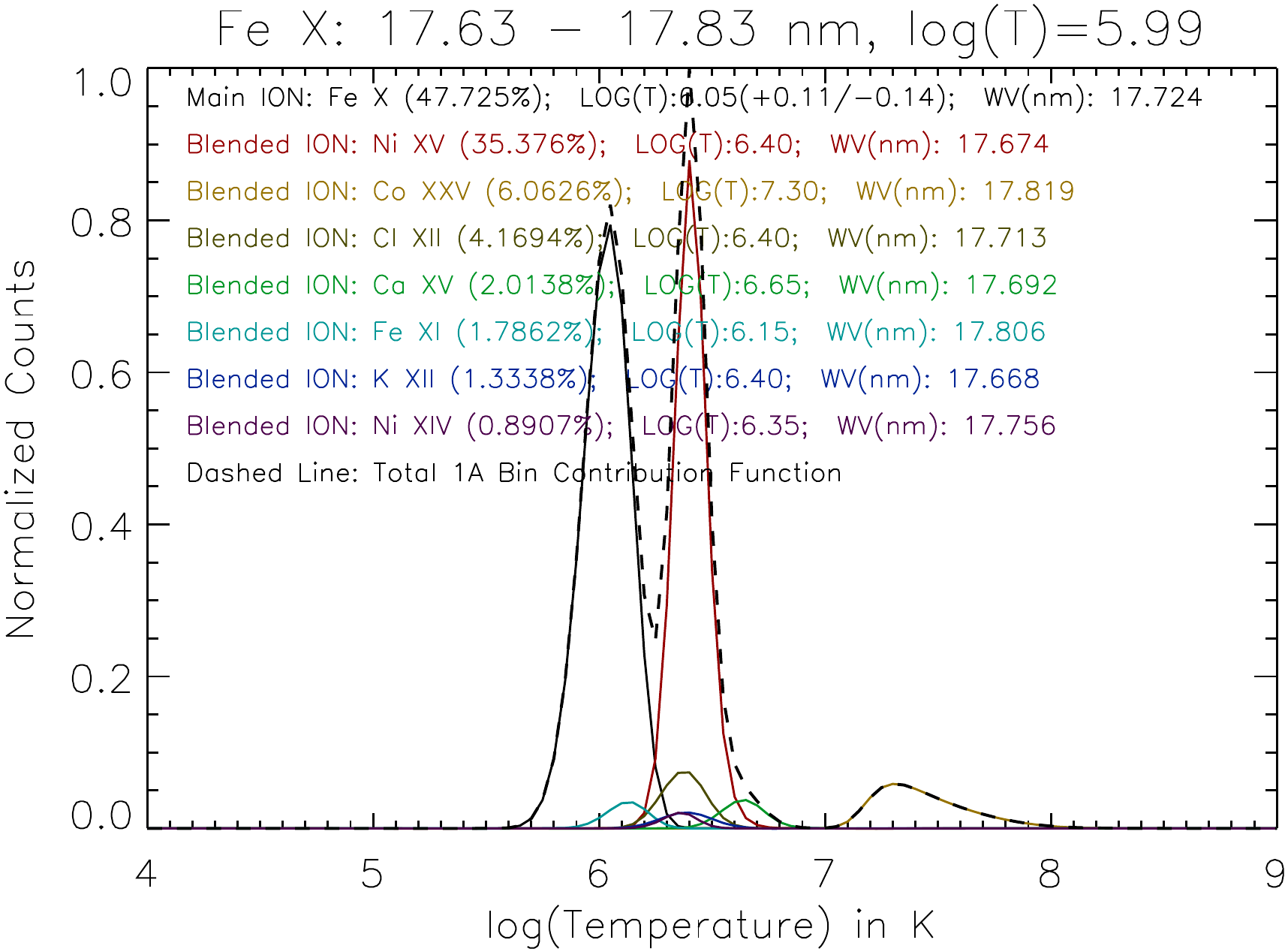}
  \caption{Temperature response curves derived based on CHIANTI atomic database. The left and middle panels are
for lines Fe XVI 33.54 nm and Fe XVI 36.08 nm, which have the same formation temperature as indicated in the
EVE level 2 data. CHIANTI calculation suggests that Fe XVI is the main ion for the emission around wavelength
band of 33.47 to 33.61 nm, but is a minor ion around wavelength band of 36.02 to 36.20 nm. The right panel
shows the temperature response curves for wavelength band of 17.63 to 17.83 nm, in which Fe X 17.72 nm is the
main emission but significantly blended by Ni XV 17.67 nm. In each panel, the percentages given in the parentheses
is the ratio of the contribution of the ion to the total emission within the given wavelength range, and the
temperature uncertainty of the main ion is read from full width at half maximum of its emission peak.}\label{fg_trc}
\end{figure}

Table~\ref{tb_lines} lists the extracted emission lines provided by MEGS-A EVL product
as well as the main temperatures they correspond to. In our final spectrum charts,
not all of the 15 emission lines are used. First of all, there are two pairs of emission lines
corresponding to the same temperature. One pair is Fe XVI 33.54 nm and Fe XVI 36.08 nm, and
the other is Fe X 17.72 nm and Mg IX 36.81 nm. For the latter pair, we exclude the line of Mg IX 36.81 nm
as the most emission lines are from Iron ions. For the former, we use the CHIANTI atomic database (version 6.0.1,
\citealt{Dere_etal_1997a, Dere_etal_2009}) to determine which one is better for our purpose.
Based on the CHIANTI database, we may estimate the temperature responses of bound-bound emission lines.
The main procedure to call CHIANTI is `\url{CH\_SYNTHETIC.PRO}' in the
solar software (SSW, \url{http://www.lmsal.com/solarsoft/}). For different features on
the Sun, the results from CHIANTI are slightly different. Here, we set
the electron number density to be $10^{11}$ cm$^{-3}$~\citep{Milligan_etal_2012a}, and assume
the abundance for solar corona and the region for
active regions, where the most flares originate. The left and middle panels of Figure~\ref{fg_trc} show the
temperature response curves of the wavelength ranges from 33.47 to 33.61 nm and from 36.02 to 36.20 nm,
where the two lines Fe XVI 33.54 nm and Fe XVI 36.08 nm located. It is obvious that within the wavelength
range of 33.47 -- 33.61 nm the emission from Fe XVI 33.54 nm is highly pronounced, whereas within the wavelength
range of 36.02 -- 36.20 nm the main ion is Mn XV. Thus, the line of Fe XVI 36.08 nm is excluded in constructing
our spectrum charts.

Second, we further remove the emission lines significantly blended with multiple
ions. As an example, the right panel of Figure~\ref{fg_trc} shows the
temperature response curve for the emission within the range of 17.63 -- 17.83 nm. Although Fe X 17.72 nm makes
the main contribution within the wavelength range, its contribution is less 48\%, and the emission from
Ni XV 17.67 nm, which corresponds to the log temperature of 6.40, is also very strong.
Here we consider an emission line being significantly blended when the contribution of the desired ion is less than 55\% of
the total emission within the wavelength range. Under this criterion, we remove the emission lines No.6, 7, 10, 14 and 15
(ref. to Table~\ref{tb_lines}), i.e., Fe XIV 21.13 nm, Fe XIII 20.20 nm, Fe X 17.72 nm, He II 25.63 nm and He II 30.38 nm,
from our TDS charts.
The final EVE TDS chart constructed based on MEGS-A data is just like that shown in Figure~\ref{fg_example_f}, in which
8 emission lines covering the logarithm of temperature from 5.57 to 6.97 are used (as marked by the asterisks in Table~\ref{tb_lines}).

\subsection{Data processing}\label{sec_data}
The procedure of generating the TDS chart consists of the following steps.

\begin{figure}[tbh]
  \centering
  \includegraphics[width=0.8\hsize]{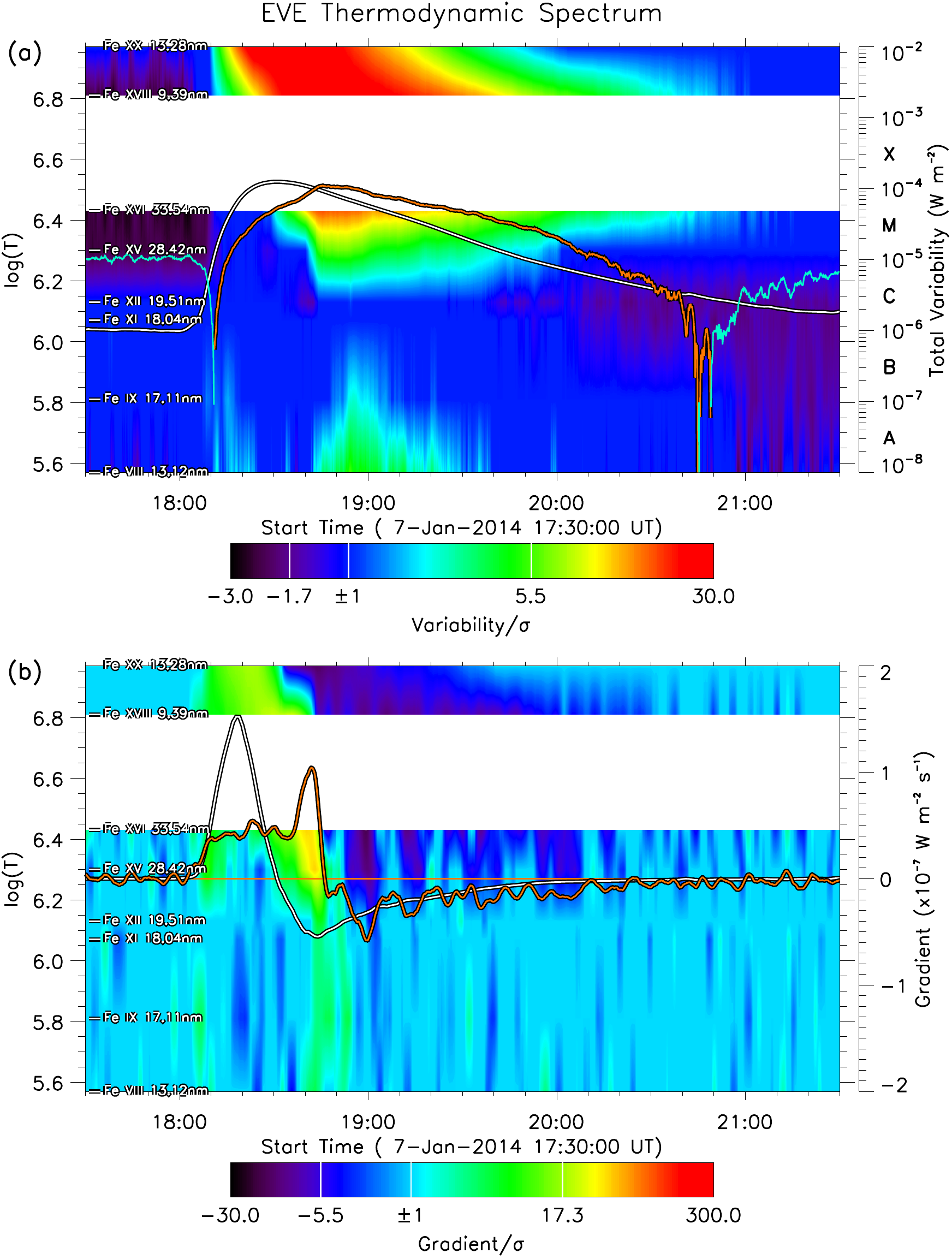}
  \caption{EVE TDS charts. Panel (a) shows the variability and Panel (b) the gradient.
See Sec.\ref{sec_data} for the definitions of the variability and gradient.
The eight emission lines used in the charts are indicated on the left. The orange line in Panel (a) is the
total variability of these lines with y-axis on the most right, and that in Panel (b) is
the gradient of the total variability. The white
lines are derived from GOES SXR for the wavelength bands of 1 -- 8 \AA.}\label{fg_example_f}
\end{figure}

\begin{figure}[tbh]
  \centering
  \includegraphics[width=\hsize]{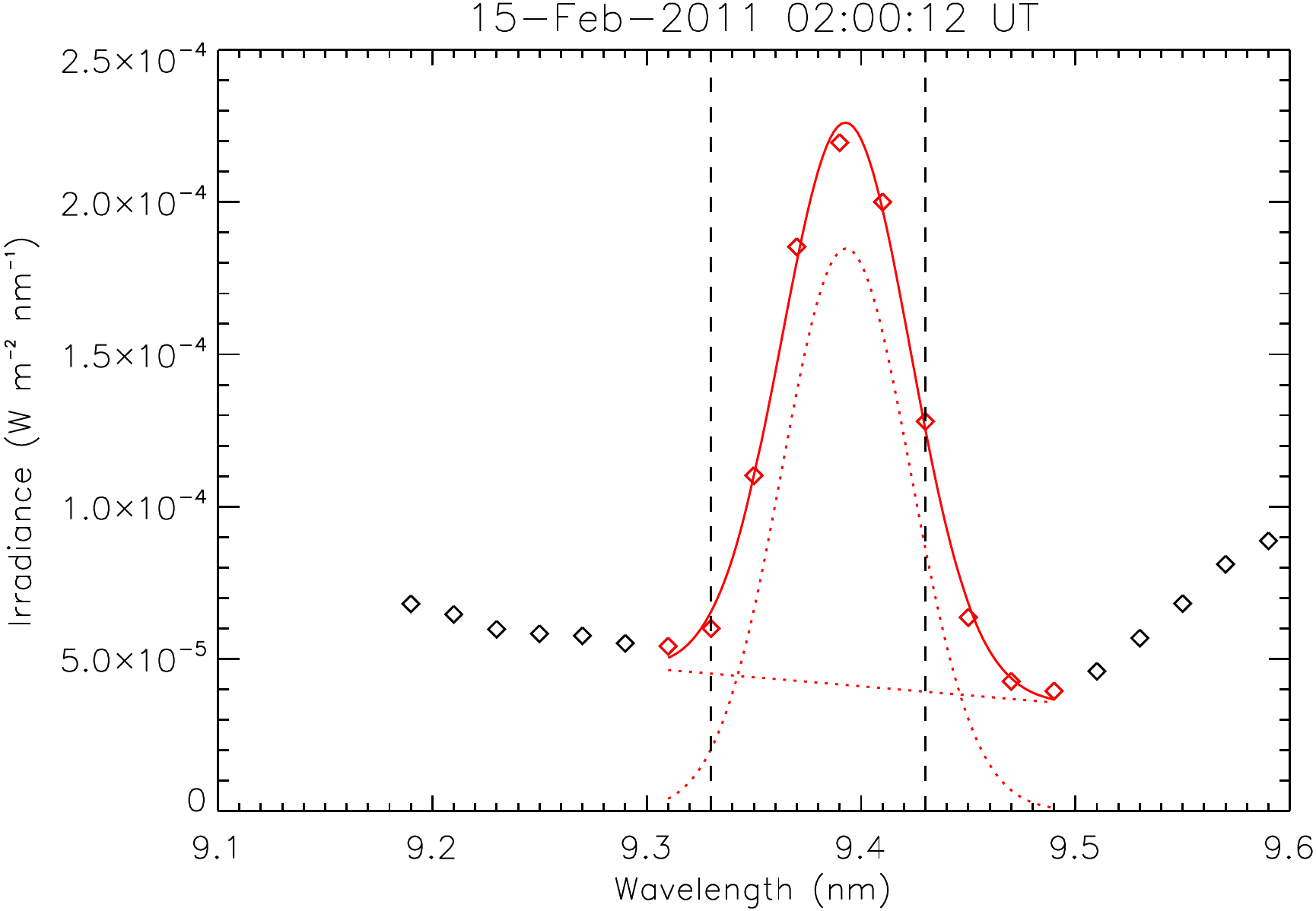}
  \caption{An example showing the EVS line profile around 9.39 nm. The two vertical
dashed lines indicate the wavelength range of the line of Fe XVIII 9.39 nm used in the EVL data. The diamonds in
red are used in our fitting procedure. The solid curve is the fitting results which consists of a Gaussian component
and a linear component as denoted by the dotted lines.}\label{fg_gaussfit}
\end{figure}

\paragraph{1. Extract emission lines.} All the 8 emission lines selected for TDS (as shown in Fig.\ref{fg_example_f})
can be found in the EVL product. In that product, however, the background continuum is not deducted. Thus,
we re-extract the lines from the EVS product. At any given time, the spectra data provide the irradiance
as a function of the wavelength. We use the information provided in the EVL data to secure the wavelength
range of each emission line of interest in the EVS data, and then use a linear combination of a Gaussian and a linear function
to fit the line profile. To avoid any possible contamination from neighbouring lines, only the data points from the
nearest local minimum on the left-hand side of the line peak to the nearest local minimum on the right-hand side of the peak
are selected for the fitting. Figure~\ref{fg_gaussfit} shows an example.
We treat the linear component of the fitting
is the background continuum at this particular time for the line of interest, and subtract it from the total irradiance
within the wavelength range of interest. This procedure is applied for all the selected emission lines all the times.
The resultant data are then input to the next step for TDS construction.

\begin{figure*}[tbh]
  \centering
  \includegraphics[width=\hsize]{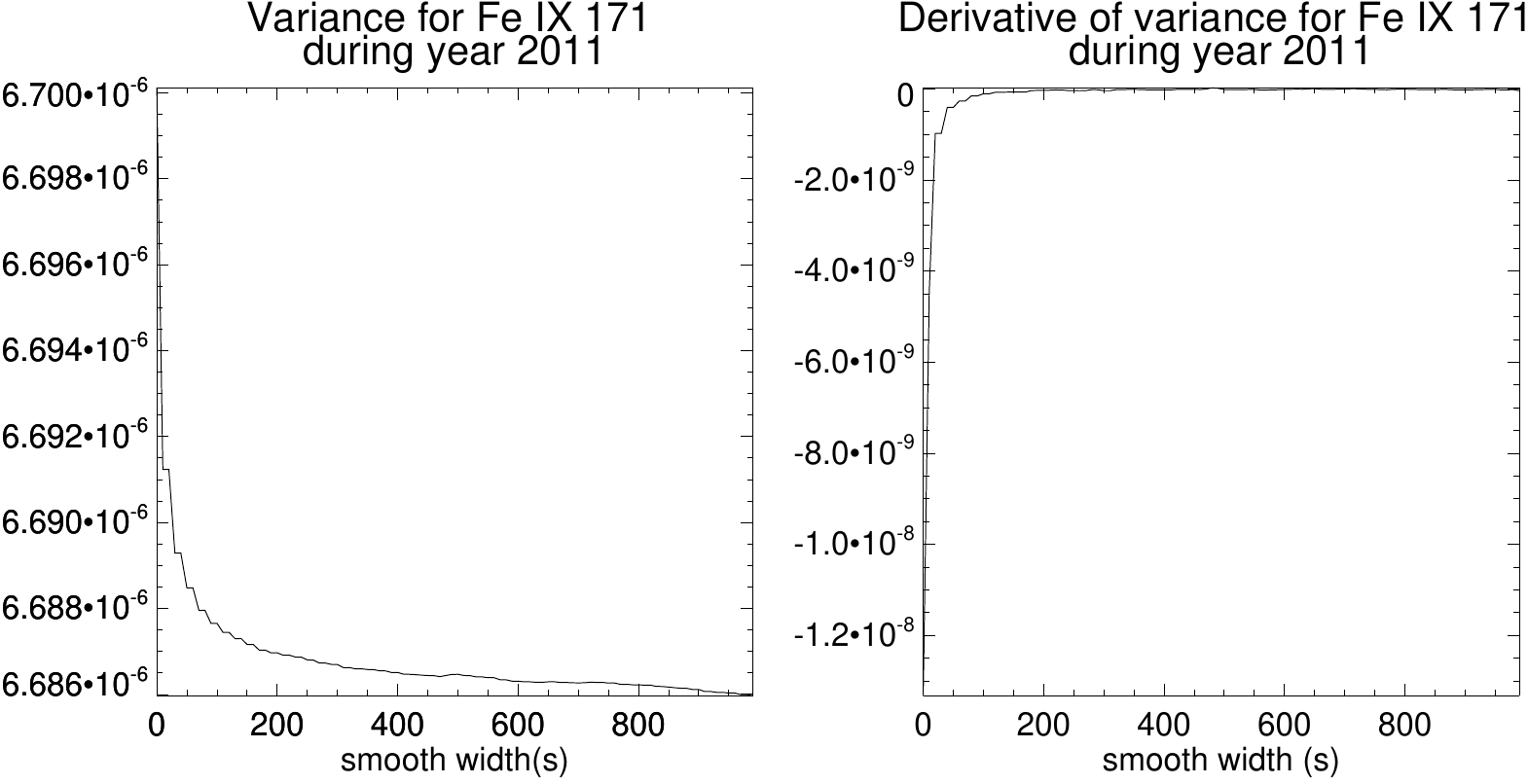}
  \caption{Dependence of the variance (left panel) of Fe IX 17.1 nm emission intensity during the
year of 2011 and its derivative (right panel) on the smooth width.}\label{fg_smooth}
\end{figure*}

\paragraph{2. Smooth data.}
The noise, regardless of the instrument noise or small irradiance variations from the full-Sun measurements,
of the EVE level 2 data is significant at the cadence of 10 seconds. Thus the second step is to
reduce the noise by smoothing data. To evaluate the level of the noise, we calculate the variance
of the data in the year of 2011. Figure~\ref{fg_smooth}, for example, shows the variance of the
emissions of Fe IX 17.11 nm as well as the derivative of the variance. From the plots, we can see that
the variance drops dramatically as the smooth width increases from 10 to 120 s, and then the drop
slows down. Thus, we choose a 2-minute time window to smooth the data. The smoothed data is labeled as
$I(t, \lambda)$, in which $t$ is the time
with a cadence of 10 second, and $\lambda$ is the wavelength of one of the eight selected emission lines.
It should be noted that, although the smoothed data still has a cadence of 10 second, some features on a timescale shorter
than 2 minutes may have been wiped away.

\begin{figure*}[tbh]
  \centering
  \includegraphics[width=\hsize]{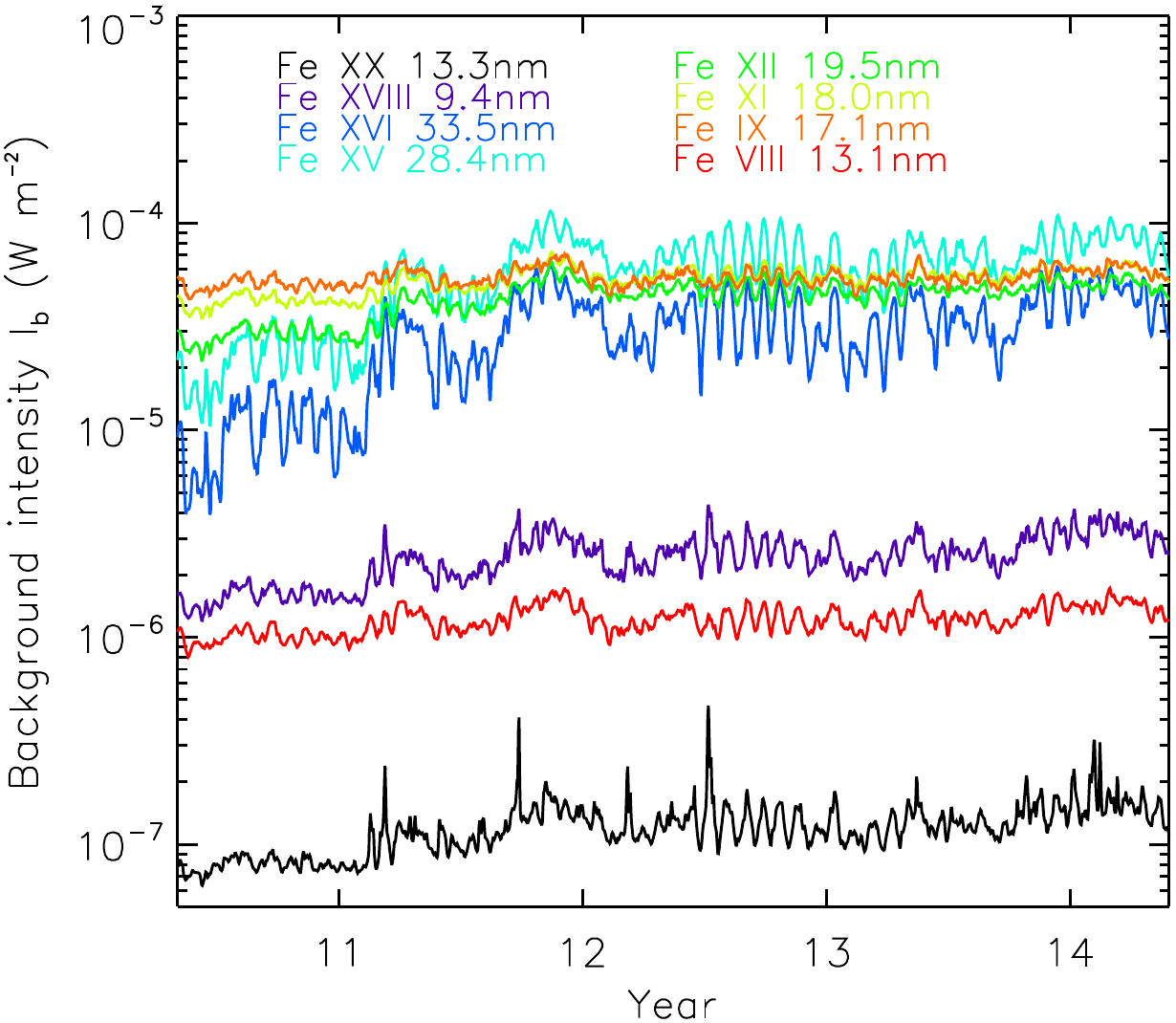}
  \caption{Background emission intensity of the eight selected emission lines.}\label{fg_bg}
\end{figure*}

\paragraph{3. Quantify solar background.} What we care about is the variability of the emission intensity during solar activities,
e.g., solar flares. To isolate the solar flare variability from the rest of the solar variations
(e.g. solar cycle, solar rotation, active region evolution, etc.), we estimate the background emission and subtract it from the smoothed data.
For a solar eruption the time scale is on the order of hours, so for any given time,
the background emission of an emission line
of interest is set to be the median value of the intensity of this line for the past 48 hours. Thus, the
background is also a function of time and wavelength, $I_b(t, \lambda)$.
Figure~\ref{fg_bg} displays the intensities of the background emissions
for reference. Obviously, this method of background emission calculation could be operated automatically and is very
useful for statistical studies. However, the obtained background emissions may sometimes be contaminated by preceding flares, particularly
when there are many flares within the 48 hours prior to the event of interest.

\paragraph{4. Calculate variability and the deviation.} The variability is defined as $I_v(t,\lambda)=I-I_b$.
It gives the intensity of an activity with the background emission subtracted. The
value of the variability could be positive, $I_{v+}$, or negative, $I_{v-}$. It is found that during the same event
the variability of different emission lines is quite different, which means that the sensitivities of the emission lines
to solar activities are different. To measure the sensitivities of the emission lines and make a uniform standard crossing
various events, we calculate the deviations of the positive and negative variabilities away from zero, respectively,
by using the whole data in 2011, i.e., $\sigma_{I\pm}(\lambda)=\sqrt{\frac{1}{N-1}\sum_tI_{v\pm}^2(t,\lambda)}$, in which
$N$ is the number of data points in time sequence. The values of $\sigma_{I\pm}$ have been listed in Table~\ref{tb_lines},
from which we can find that the lines
Fe XV 28.42 nm and Fe XVI 33.54 nm are most sensitive to solar activities among the 8 emission lines, and
line Fe VIII 13.12 nm is the most insensitive.

By normalizing the variability, $I_{v\pm}$, by the deviation, $\sigma_{I\pm}$, different emission lines can be
compared. We plot the normalized variability in the time-temperature plane to generate the TDS chart,
as shown in the upper panel of Figure~\ref{fg_example_f}. The small gaps between
the selected lines (or temperatures), except the large one between $\log(T)$ of 6.81 and 6.43, are simply filled by applying
linear interpolation. The total variability, $\sum_\lambda I_v(t,\lambda)$, is superimposed on the chart as the orange line
(Note, due to the logarithmic scale, for the negative total variability, a cyan line is used).
For comparison, the GOES SXR flux from the wavelength band of 1 -- 8 \AA\ is superimposed as the white line.

A similar procedure is used to generate the spectrum chart of the gradient of the line intensity.
Based on the 2-min smoothed data, we derive the gradient, $G_{\pm}(t, \lambda)=\frac{dI(t,\lambda)}{dt}$, by linearly fitting
the intensity, $I(t, \lambda)$, within a time window of 5 minutes. Then we calculate the deviations of
the gradients away from zero by using the whole data in 2011, i.e., $\sigma_{G\pm}(\lambda)=\sqrt{\frac{1}{N-1}\sum_tG_{\pm}^2(t,\lambda)}$,
and plot the chart in the logarithm of $\frac{G_{\pm}}{\sigma_{G\pm}}$, as shown in the
lower panel of Figure~\ref{fg_example_f}. The gradient of the total variability is indicated by the orange line,
and the gradient of the GOES SXR flux by the white line.

An online website has been established to exhibit the TDS charts \\
(\url{http://space.ustc.edu.cn/dreams/shm/tds}).
From the TDS charts, we can learn the start and end times of the
enhanced/reduced emissions, the temperatures at which the emissions enhance/reduce, drift
rate of the temperature, rising and declining rates of the enhancement/reduction, etc. Particularly,
the relative variability, i.e., the radiative output with the background deducted, of EUV emission provides
us information to reveal the plasma thermodynamics in the middle to high corona, and may be also useful
in studying the change of Earth's atmosphere as well as its associated physics mechanisms which is partially
driven by, e.g., solar flares~\citep[e.g.,][]{Sutton_etal_2006, Pawlowski_Ridley_2008, Qian_etal_2010}.

\section{Four different types of flares}\label{sec_cases}
Traditionally, flares are classified as confined and eruptive flares. The former is not
associated with a coronal mass ejection (CME) and the latter is. Recently, SDO observations reveal that flares may
not only have one main phase of emission, but also experience a late phase, i.e., there is
a second peak of the flare emission \citep{Woods_etal_2011}.
Combining the two different features, we may classify the flares as confined/eruptive flares
with/without a late phase. In the following sections, we will present four M-class flares
in these four different types. The four flares except the first one were all studied before as listed
in Table~\ref{tb_flares}. By investigating these flares, we justify our method and
also show the flare signatures in the TDS charts.

\subsection{A confined flare on 2011 September 8}
\begin{figure*}[tbh]
  \centering
  \includegraphics[width=\hsize]{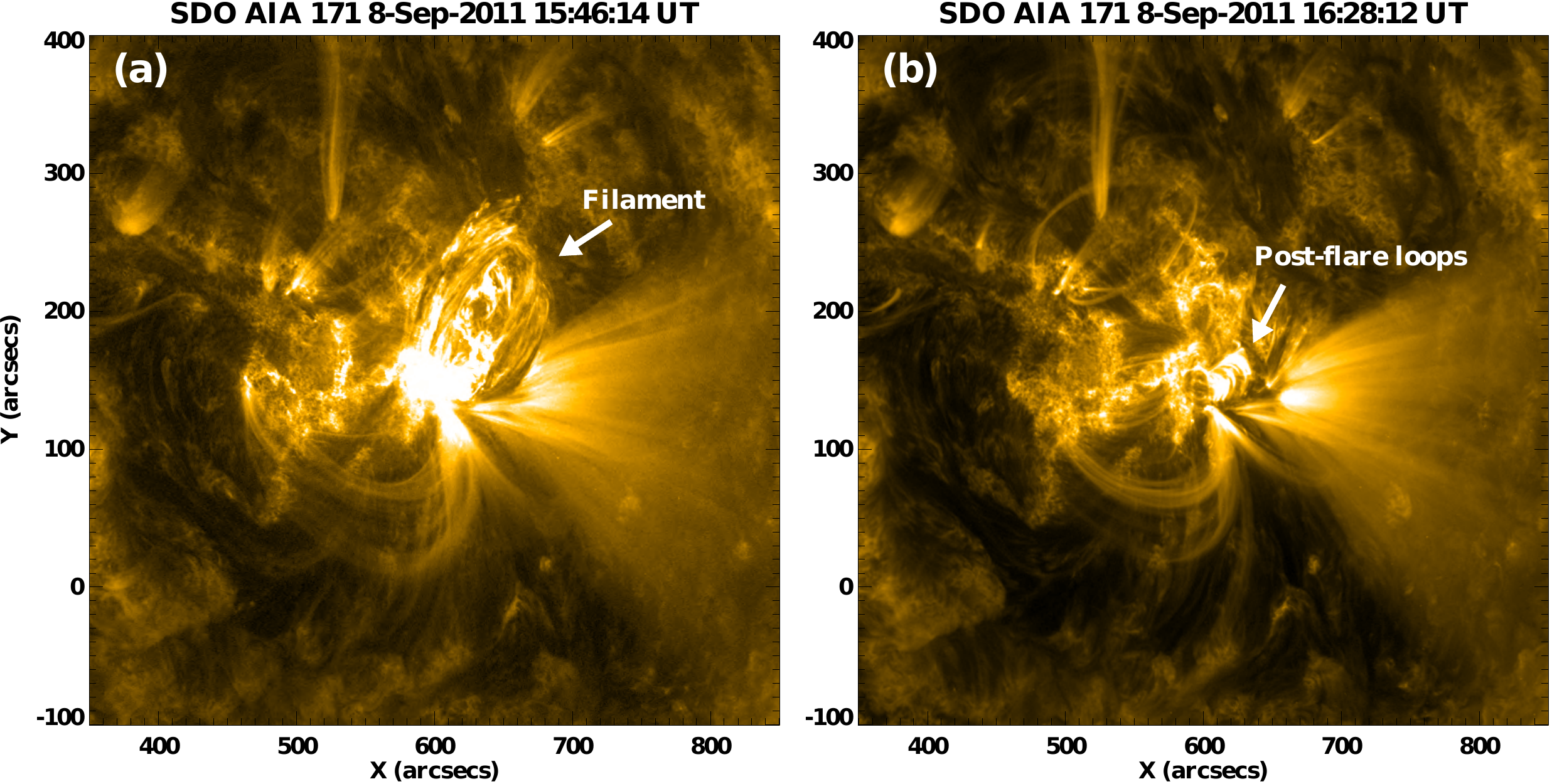}
  \caption{SDO/AIA 171 images of the 2011 September 8 flare (Case C1) at and after the peak time.}\label{fg_case_c0}
\end{figure*}

Based on the GOES SXR flux, the 2011 September 8 flare is an M6.7 X-ray flare, starting at 15:32 UT and peaking at 15:46 UT, and
the whole duration of the flare is 20 minutes.
Figure~\ref{fg_case_c0}a and \ref{fg_case_c0}b show the SDO/AIA 171 images at and after the peak of the flare.
It happened in a compact region on the west hemisphere. An active-region filament rose during the flare
but did not erupt out, and post-flare loops were clear. There
was no dimming in the EUV images and no CME observed by SOHO/LASCO \citep{Brueckner_etal_1995}, suggesting a confined flare.

\begin{figure}[tbh]
  \centering
  \includegraphics[width=0.8\hsize]{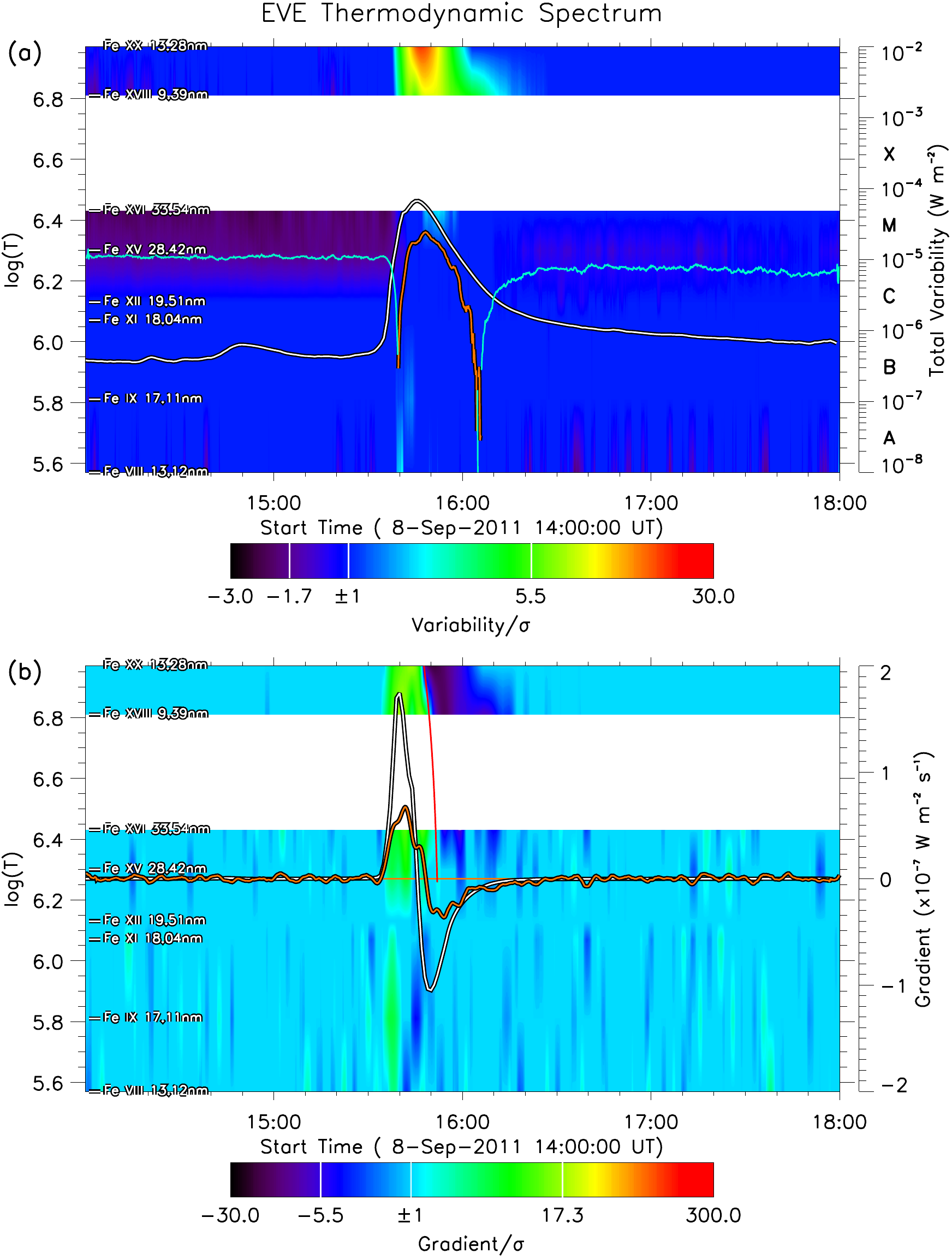}
  \caption{EVE TDS charts for Case C1, the 2011 September 8 flare. All the decorations have
the same meanings of those in Fig.\ref{fg_example_f} except the red line in Panel (b), which is given by
the fitting of the linear cooling. The cyan curve of the total variability in Panel (a) means a dimming, i.e.,
the variability is less than zero.}\label{fg_tds_c0}
\end{figure}

In the EVE TDS chart, there were clear enhancements during
the flaring period as shown in Figure~\ref{fg_tds_c0}a and \ref{fg_tds_c0}b. During the same period, there was no other flare on
the visible solar disk, and therefore the features displayed in the spectrum chart reflect the thermal
processes of the flare.
First, the flare heated the coronal lines simultaneously, and the emission enhancements of the lines
at the high temperatures were more significant than those at the low temperatures.
The start time is defined by the significant deviation from zero of the orange curve
in the gradient chart (Fig.\ref{fg_tds_c0}b), which is around 15:32 UT, the same as reported for GOES SXR. The peak time is about
15:48 UT, 2 minutes later than that of GOES SXR (reading from the curves in Fig.\ref{fg_tds_c0}a),
reflecting the time scale of the cooling process of extremely hot X-ray emission plasma to less-hot EUV plasma.
For a GOES X-ray flare, people traditionally defines the end time when the current SXR flux returns to half of the peak
flux\footnote{refer to \url{http://www.ngdc.noaa.gov/stp/solar/solarflares.html}}.
Here we would like to define the end time of a flare as the time when the gradient indicated by the orange curve
in Fig.\ref{fg_tds_c0}b returns back to zero. Due to the different definitions, our estimated duration of the flare,
which is about 40 minutes, is much longer than that from the GOES report.

Second, the cooling process of the heated thermal plasmas at high temperatures are notable in both variability
and gradient charts. Particularly, the cooling is clearly revealed by the drift of the
interface between the positive and negative gradients as shown in the gradient chart
(Fig.\ref{fg_tds_c0}b). The drift rate characterizes the overall cooling rate of the flare plasma that
is a combined effort of radiative cooling and conductive cooling. By using the linear function, i.e., $T=T_0+c_rt$,
to fit the interface, we can estimate the linear cooling rate, $c_r$, is
about $-0.03\pm0.01$ MK s$^{-1}$ for the heated plasma. Here when doing fitting, we set the uncertainty
in temperature to be $\pm0.15$ in $\log(T)$,
which is approximately the full width at half maximum of the main peak of the temperature response curve for all the 
selected emission lines.
The red curve in Figure~\ref{fg_tds_c0}b show the fitting result.
It should be noted that the estimated cooling
rate is a lower limit because most flares continuously release magnetic energy
and heat the plasmas throughout the entire phase \citep[e.g.,][]{Jiang_etal_2006, Warren_2006, Ryan_etal_2013}.

There was no late phase associated with this flare. Besides, one may notice that the variability chart shows
a significant dimming, i.e., the decrease of the emissions, before the flare
(the dark region near $\log(T)$ of 6.2 to 6.4 in Fig.\ref{fg_tds_c0}a).
It is not caused by any solar activity, but is the consequence of a bright active region
on the west limb rotating off from the visible solar disk.

\subsection{A confined flare with a late phase on 2010 November 5}
\begin{figure*}[tbh]
  \centering
  \includegraphics[width=\hsize]{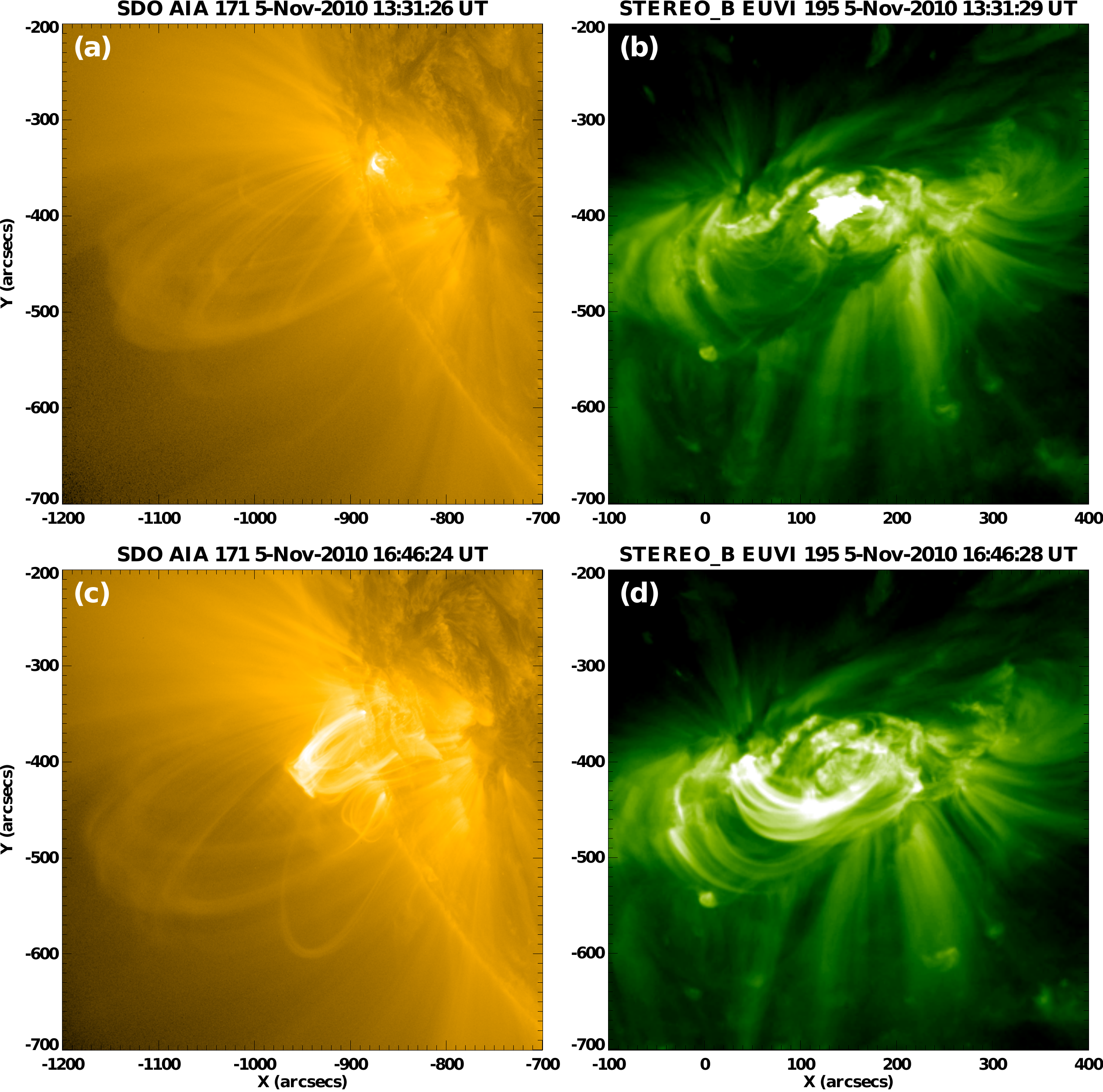}
  \caption{The upper panels show the side view and top view of the 2010 November 5 flare
near the peak time of the main phase from the SDO/AIA 171 (the left panel) and STEREO-B/EUVI 195 (the right panel).
The lower panels are the snapshots near the peak time of the late phase.}\label{fg_case_c1}
\end{figure*}

This flare was studied by \citet{Woods_etal_2011}, \citet{Chamberlin_etal_2012} and \citet{LiuK_etal_2015}.
It started at 12:43 UT, peaked at 13:29 UT, and lasted 83 minutes according to the GOES report.
Different from the previous one, the flare has a much longer decay phase. Its main phase
occurred in a compact region (Fig.~\ref{fg_case_c1}a and \ref{fg_case_c1}b), but the late phase was due to the
brightening of the neighboring loops in a larger region (Fig.~\ref{fg_case_c1}c and \ref{fg_case_c1}d).
Although the flare is as intense as M1.0 and is long lasting, no CME was associated.

\begin{figure}[tbh]
  \centering
  \includegraphics[width=\hsize]{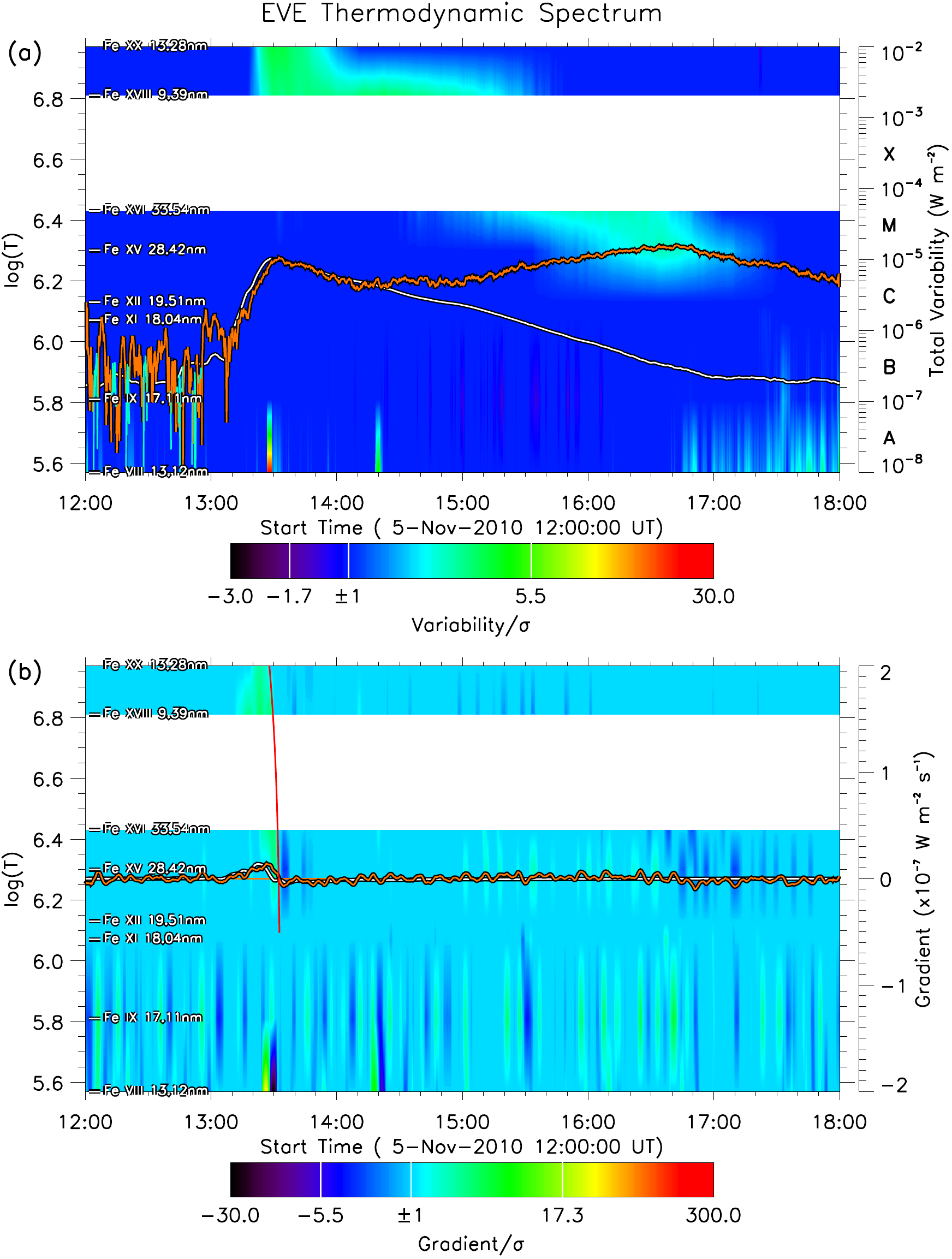}
  \caption{EVE TDS charts for Case C2, the 2010 November 5 flare.}\label{fg_tds_c1}
\end{figure}

The main and late phases of the flare are clearly shown by the total variability in
Figure~\ref{fg_tds_c1}a. The peak of the late phase is higher than that of the main phase, suggesting that
additional magnetic and/or kinetic energies were converted into thermal energy
during the late phase, which was even larger than that during the main phase. However,
the GOES SXR did not show any signature of the late phase, further implying that the energy
conversion during the late phase was probably through a much gentler way than that during the main phase.

Based on the spectrum charts, the flare began at about 13:15 UT and ended after 18:30 UT (exceeding
the time range of the charts), and the first and second peaks occurred around 13:32 and 16:42 UT, respectively.
Compared with the GOES SXR, the peak time of the main phase is about 3 minutes late, which is similar to
the previous case, and the duration of the flare in EUV passbands is much longer than that in SXR. Such a
long-duration confined flare is contrary to the traditional picture that long-duration flares tend to be
eruptive \citep[e.g.,][]{Harrison_1995, Yashiro_etal_2006}, implying a strong constraint above the
flare region \citep[e.g.,][]{Wang_Zhang_2007, LiuY_2008}. Another case could be found in the paper
by~\citet{LiuR_etal_2014}, who reported a long-duration confined X-class flare.

At the beginning of the flare, the plasma was mainly heated at a high temperature above 6.3 MK, and then
the enhancement apparently drifted down to around 2.0 MK when the second peak occurred. Since the
enhanced emissions in the main and late phases came from the different regions as indicated in
Figure~\ref{fg_case_c1}a and \ref{fg_case_c1}b, the drift feature in Figure~\ref{fg_tds_c1}a
cannot be interpreted as a coherent cooling process. Actually, it is a combination of a
cooling process during the main phase and an additional heating and cooling process during the late phase.
The cooling signature during the main phase could be clearly recognized in the gradient chart
(as indicated by the red linear fitting line in Fig.\ref{fg_tds_c1}b), from which
the linear cooling rate is estimated as about $-0.03\pm0.01$ MK s$^{-1}$.

\subsection{An eruptive flare on 2011 March 8}
\begin{figure*}[tbh]
  \centering
  \includegraphics[width=\hsize]{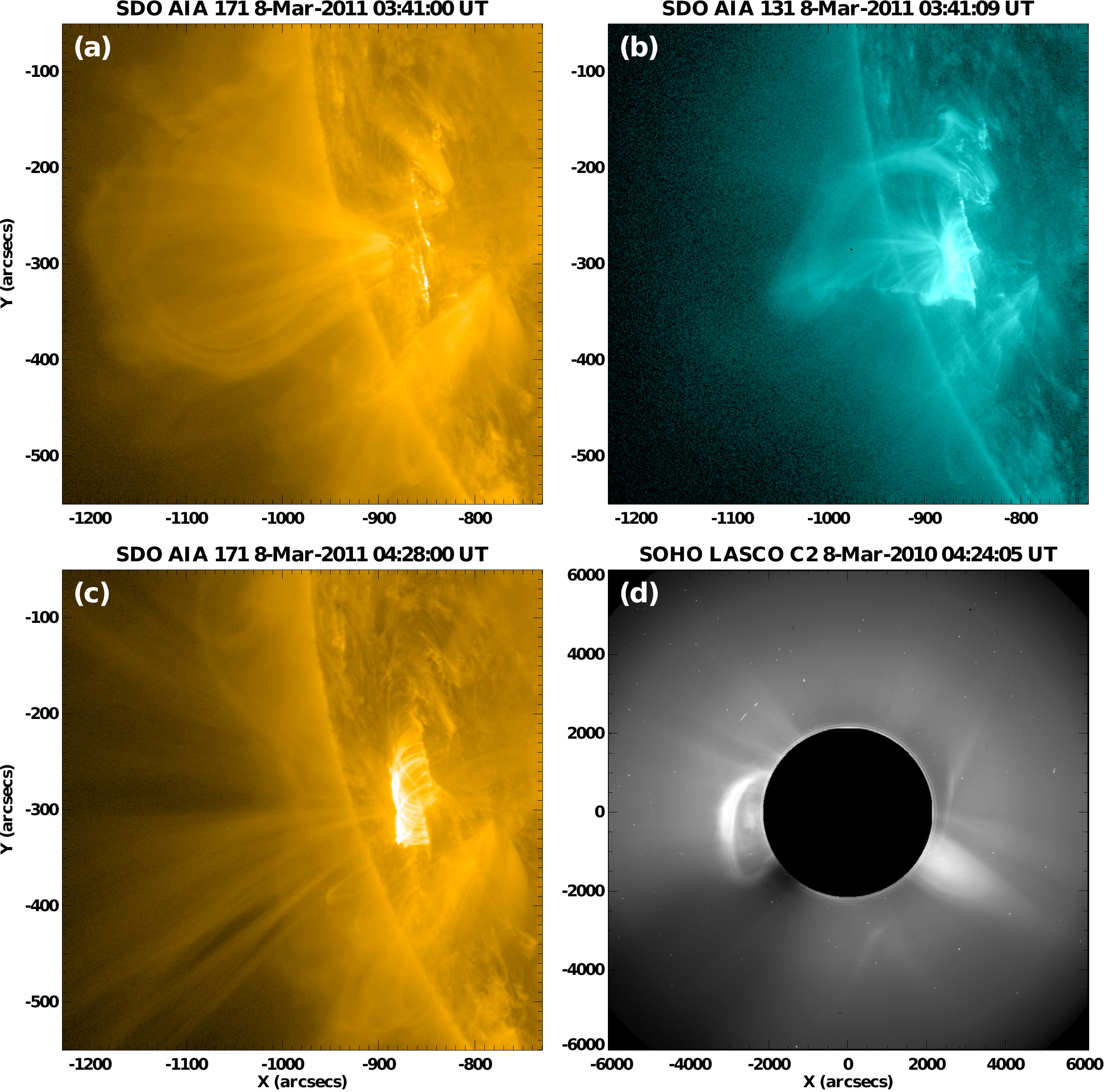}
  \caption{The upper panels show the arcades viewed in SDO/AIA 171 (the left panel) and the underneath flux rope viewed
in SDO/AIA 131 (the right panel) during the rising phase of the 2011 March 8 flare. Panel (c) shows the
post-flare loops after the peak time. Panel (d) displays the associated CME observed by SOHO/LASCO C2.}\label{fg_case_e0}
\end{figure*}

The flare, which started at 03:37 UT and peaked at 03:58 UT on 2011 March 8, was associated with
an eruption of a flux rope \citep{Zhang_etal_2012}. The flux rope and the overlying arcades can
been seen in the AIA 131 (Fig.\ref{fg_case_e0}b) and AIA 171 (Fig.\ref{fg_case_e0}a)
images, respectively. The flux rope developed into a CME which was recorded by
the SOHO/LASCO C2 camera as shown in Figure~\ref{fg_case_e0}d.
The post-flare loops are clear visible after the flux rope erupted (Fig.\ref{fg_case_e0}c).

\begin{figure}[tbh]
  \centering
  \includegraphics[width=\hsize]{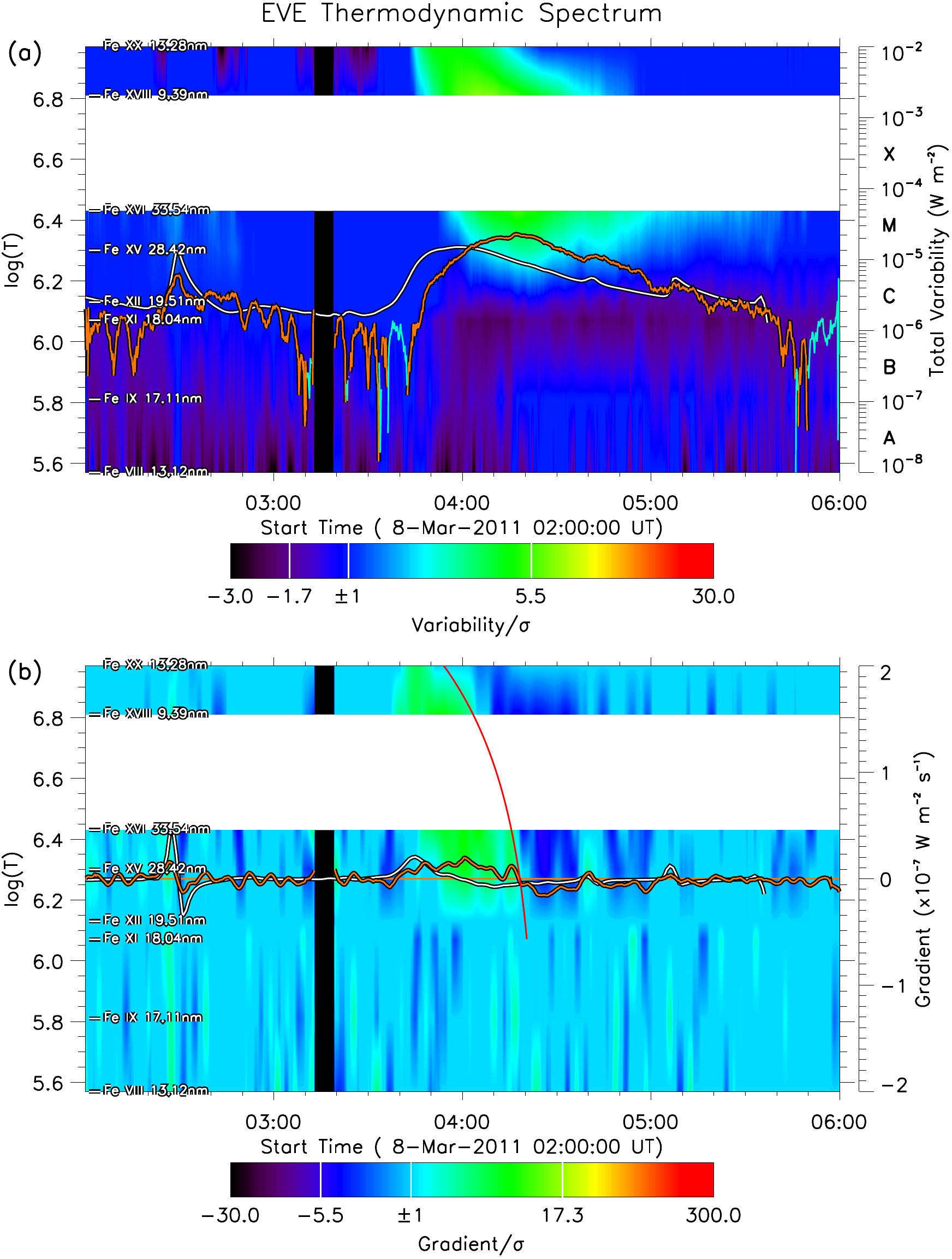}
  \caption{EVE TDS charts for Case E1, the 2011 March 8 flare.}\label{fg_tds_e0}
\end{figure}

Based on the GOES SXR, it is an M1.5 flare and the duration is about 48 minutes. On the other hand,
as we can see in Figure~\ref{fg_tds_e0}a,
the whole profile of the variability curve of the flare lags about 10 to 30 minutes behind
the SXR curve. The flare began at about 03:40 UT, peaked
at 04:17 UT and ended after 05:20 UT in EUV passbands. The EUV peak is about 19 minutes later than the
SXR peak. The delay is much longer than those in other cases, but is consistent with the slow cooling rate
of the heated plasma during the flare as will be seen below. Besides,
under our definition, the duration of the flare is more than 100 minutes in EUV, suggesting a long-duration flare.

Different from previous cases, the enhancement of the EUV emission appeared earlier at the
higher temperature, and a clear drift of the enhancement, which forms a flag-shape, could be
found in Figure~\ref{fg_tds_e0}a. There was no significant enhancement at the low temperature.
By using the gradient chart (Fig.\ref{fg_tds_e0}b), we get that the linear cooling rate of the
heated plasma is about $-0.005\pm0.002$ MK s$^{-1}$, about one order lower than the two previous cases.

Besides, the variability chart (Fig.\ref{fg_case_e0}a) suggests that
there are significant dimmings before and during the flare below the temperature of $\log(T)<6.2$.
These dimmings are all probably due to the depletion of the coronal density caused by eruptions. The dimming
before the flare is associated with an M3.7 flare peaking at 20:01 UT on the previous day.

\subsection{An eruptive flare with a late phase on 2010 October 16}
\begin{figure*}[tbh]
  \centering
  \includegraphics[width=\hsize]{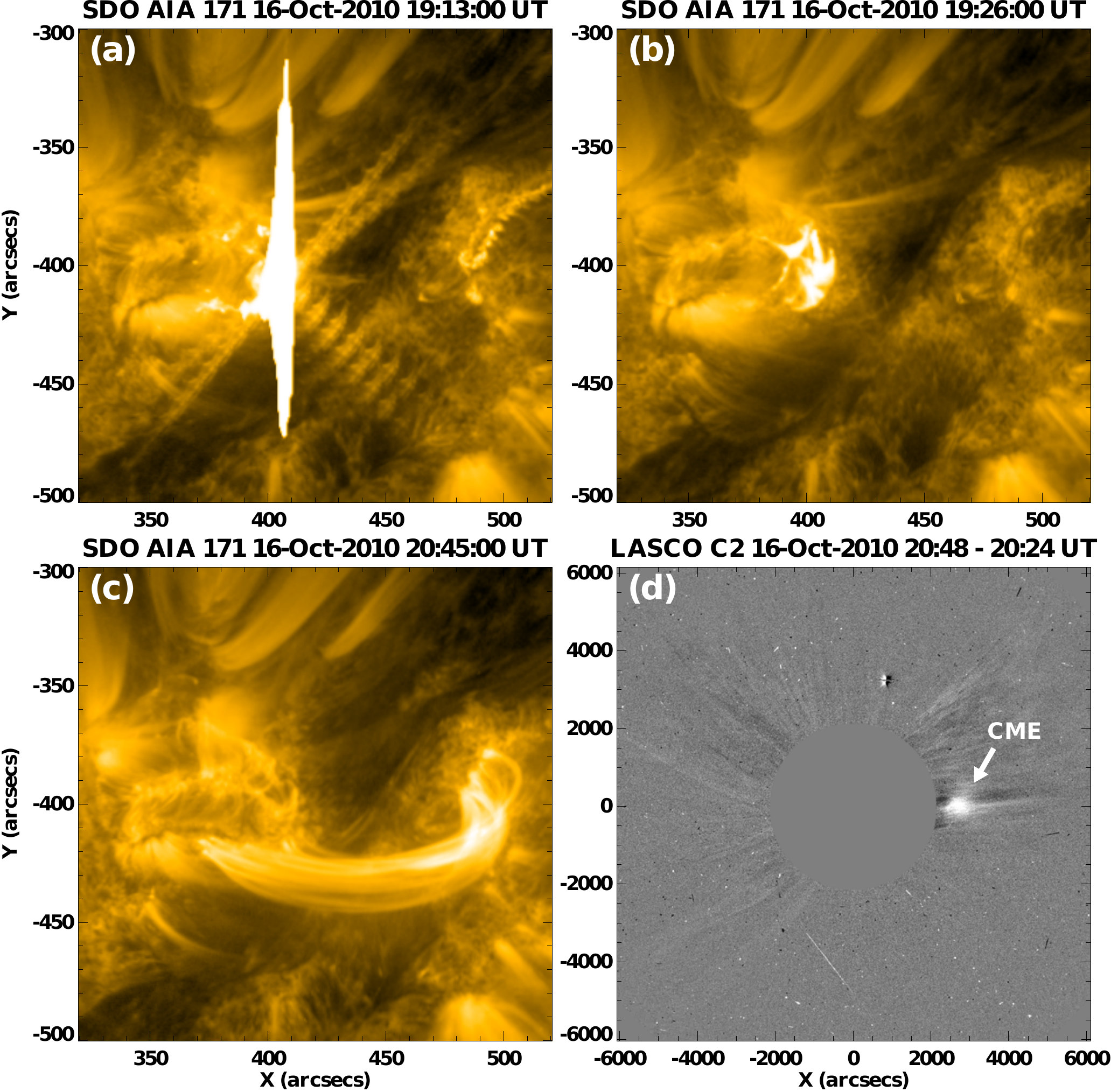}
  \caption{Panel (a)--(b) show the peak time, decline phase and late phase of the 2010 October 16 flare
viewed by SDO/AIA 171. Panel (d) displays the associated weak CME observed by SOHO/LASCO C2.}\label{fg_case_e1}
\end{figure*}

This is an impulsive M3.0 flare based on the report of GOES SXR. It started at 19:07 UT and peaked at 19:12 UT
(Fig.\ref{fg_case_e1}a). The end time in GOES definition is at the same minute as the peak time,
so that the duration of the flare is only 5 minutes. But this flare
has a late phase as suggested by \citet{Woods_etal_2011} and \citet{LiuK_etal_2013}. The enhancement of the emission
during the late phase is attributed to the neighboring loops as shown in Figure~\ref{fg_case_e1}c. This flare was
accompanied by a weak CME with post-flare loops clearly visible in the source region
(Fig.\ref{fg_case_e1}c and \ref{fg_case_e1}d).

\begin{figure}[tbh]
  \centering
  \includegraphics[width=\hsize]{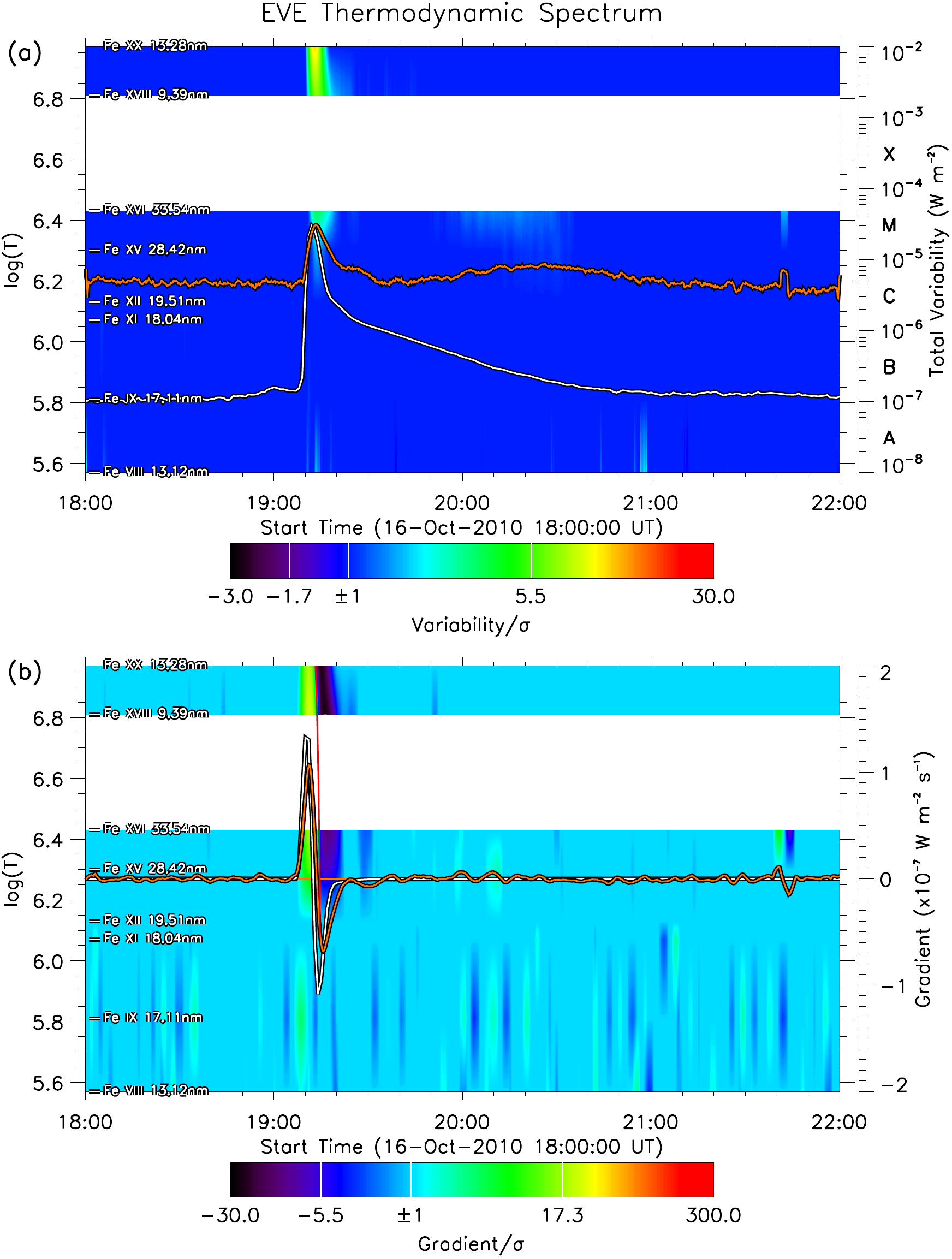}
  \caption{EVE TDS charts for Case E2, the 2010 October 16 flare.}\label{fg_tds_e1}
\end{figure}

The main and late phases of the flare can be clearly seen in Figure~\ref{fg_tds_e1}a. The main peak occurred at 19:13 UT
and the late-phase peak at 20:26 UT. Compared with the SXR, the main peak in EUV is about 1 minutes late. The start time
of the flare in EUV is about 19:09 UT and the end time is after 21:00 UT, which suggest that the flare actually lasted
much longer in EUV than in SXR.

From the gradient chart (Fig.\ref{fg_tds_e1}b), the drift of the enhancement feature from the high temperature 
to the low temperature is quite fast. Our fitting suggests that the
linear cooling rate is about $-0.10\pm0.04$ MK s$^{-1}$.
Besides of the apparent drift feature in the main phase, we also can find a very faint
drift feature from the main phase to the late phase in the variability chart
(see Fig.\ref{fg_tds_e1}a). As we pointed out before, this drift
may not be interpreted merely as a cooling process, but is a manifestation of the additional
heating of the plasma in the neighboring loops as shown in Figure~\ref{fg_case_e1}c (also refer to \citealt{LiuK_etal_2013}).

\section{Statistical results about flares equal to or greater than M5.0}\label{sec_results}

\subsection{Delay of EUV peaks and the cooling rate}\label{sec_cr}

The previous section has presented what we can learn from the TDS charts through the investigation of four
flares of different types. Here we apply our method to all the flares equal to or greater than M5.0-class.
According to GOES SXR records, there were about 75 M5.0+ flares during the EVE/MEGS-A's 5-year lifetime.
All of these flares have EVE observations except one. Table~\ref{tb_m5_flares} list the information of these flares.

The observed linear cooling rate of the heated plasma for the cases in the previous section
varies from $-0.003$ to $-0.14$ MK s$^{-1}$. Its value is roughly correlated to the delay of the peaks between the
SXR and the EUV emissions. A slower cooling rate tends to have a longer delay time, suggesting a systematic
cooling process from tens to a few million Kelvin. However, there were only four data points. To solidify the correlation,
we measure the cooling rates of all the M5.0+ flares as well as the delay times of the EUV peaks.

\begin{figure*}[tb]
  \centering
  \includegraphics[width=0.49\hsize]{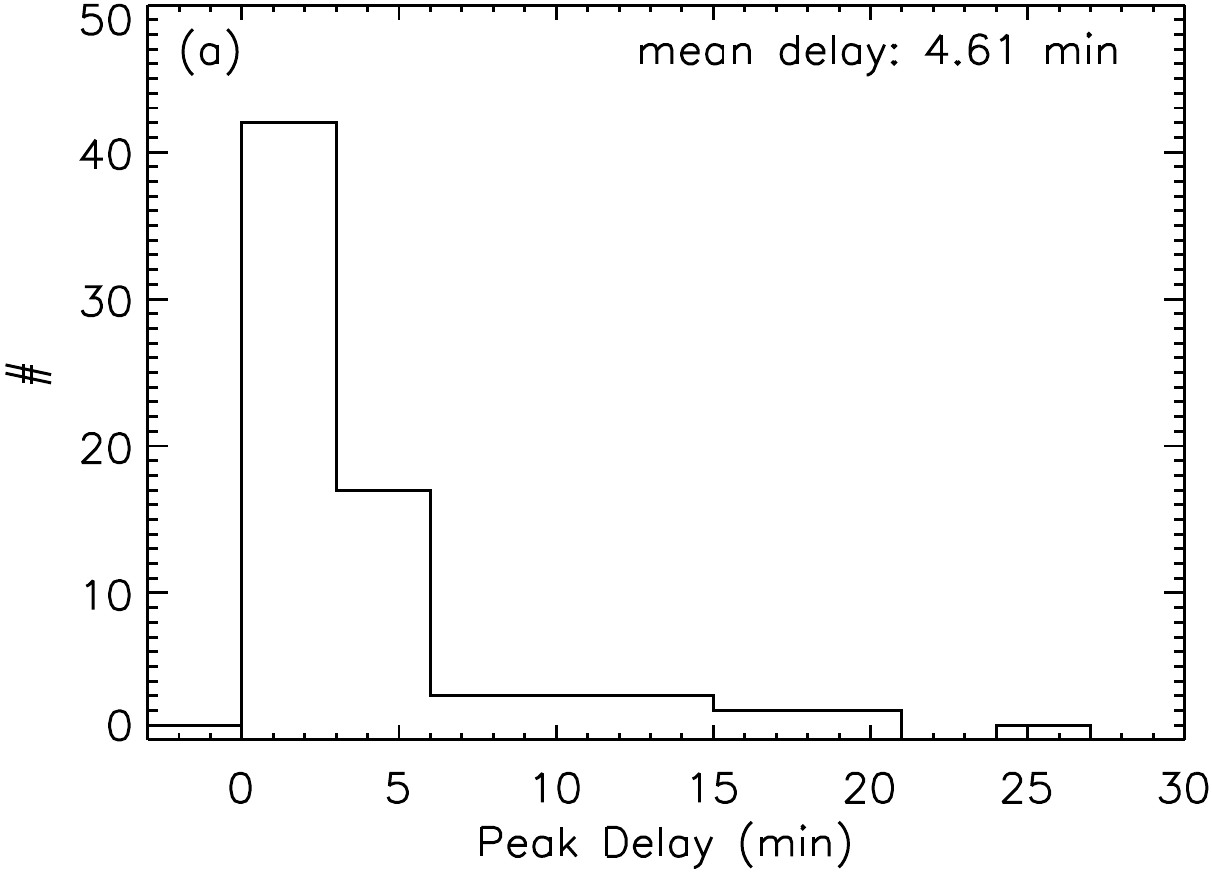}
  \includegraphics[width=0.50\hsize]{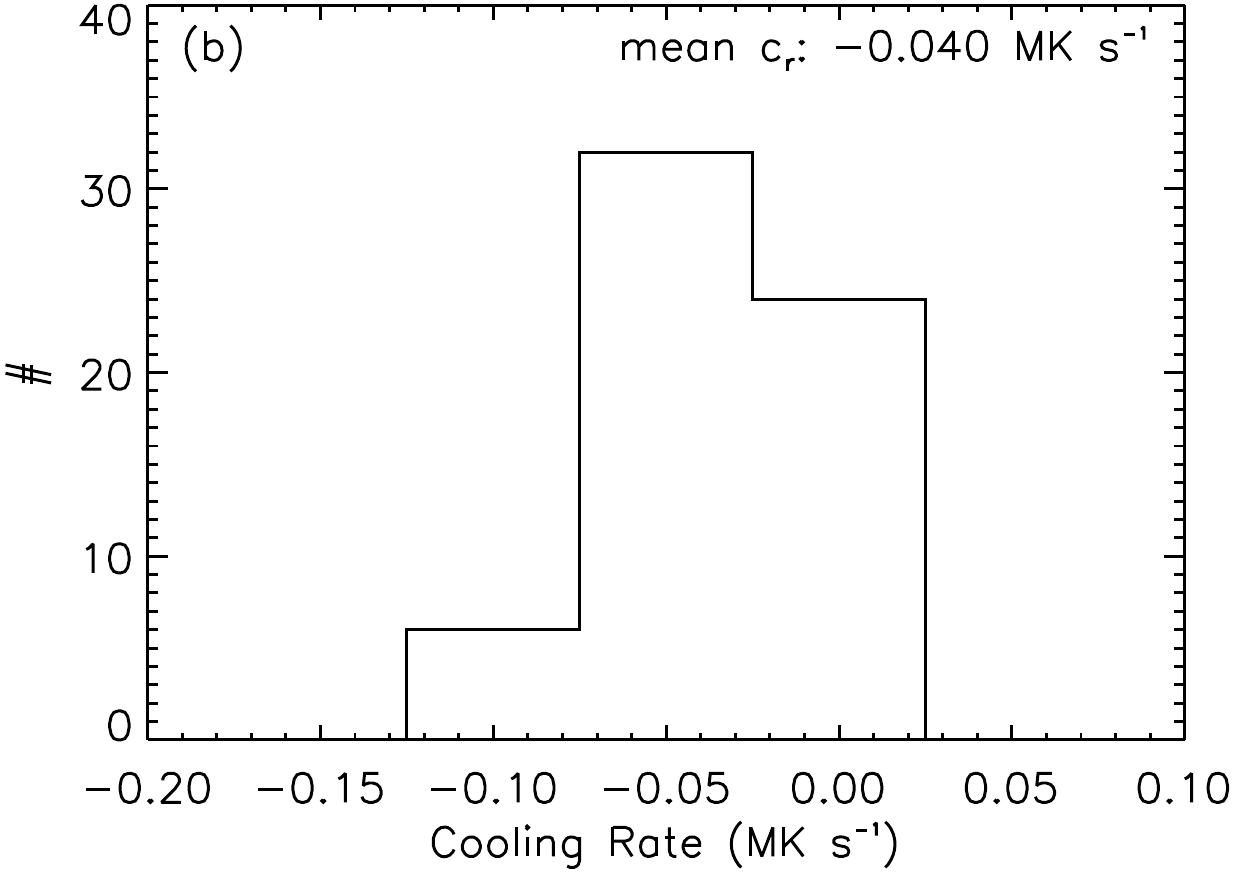}
  \caption{Histograms of the delay times of the EUV peaks with respect to the SXR peaks and the linear cooling rates measured from EVE TDS charts.}\label{fg_cr_pd_dist}
\end{figure*}

First, it is found that the EUV peak is behind of the SXR peak for all the flares except one, 
the 2012 July 19 flare (No.35 in Table~\ref{tb_m5_flares}), of which the EUV peak is 0.9-minute ahead of the SXR peak.
But for this event, it does not mean that the EUV emission reaches the peak before the SXR emission, 
because the time difference is less than one minute, which falls in the uncertainty of the data; 
our EVE data were smoothed by a 2-min time window and the cadence of the SXR data used here is one minute.
The distribution of the delay times in Figure~\ref{fg_cr_pd_dist}a shows that the delay for most flares is 
less than 6 minutes and occasionally longer than 20 minutes, and the mean value of the peak delay is about 5 minutes.

\begin{figure*}[tb]
  \centering
  \includegraphics[width=0.505\hsize]{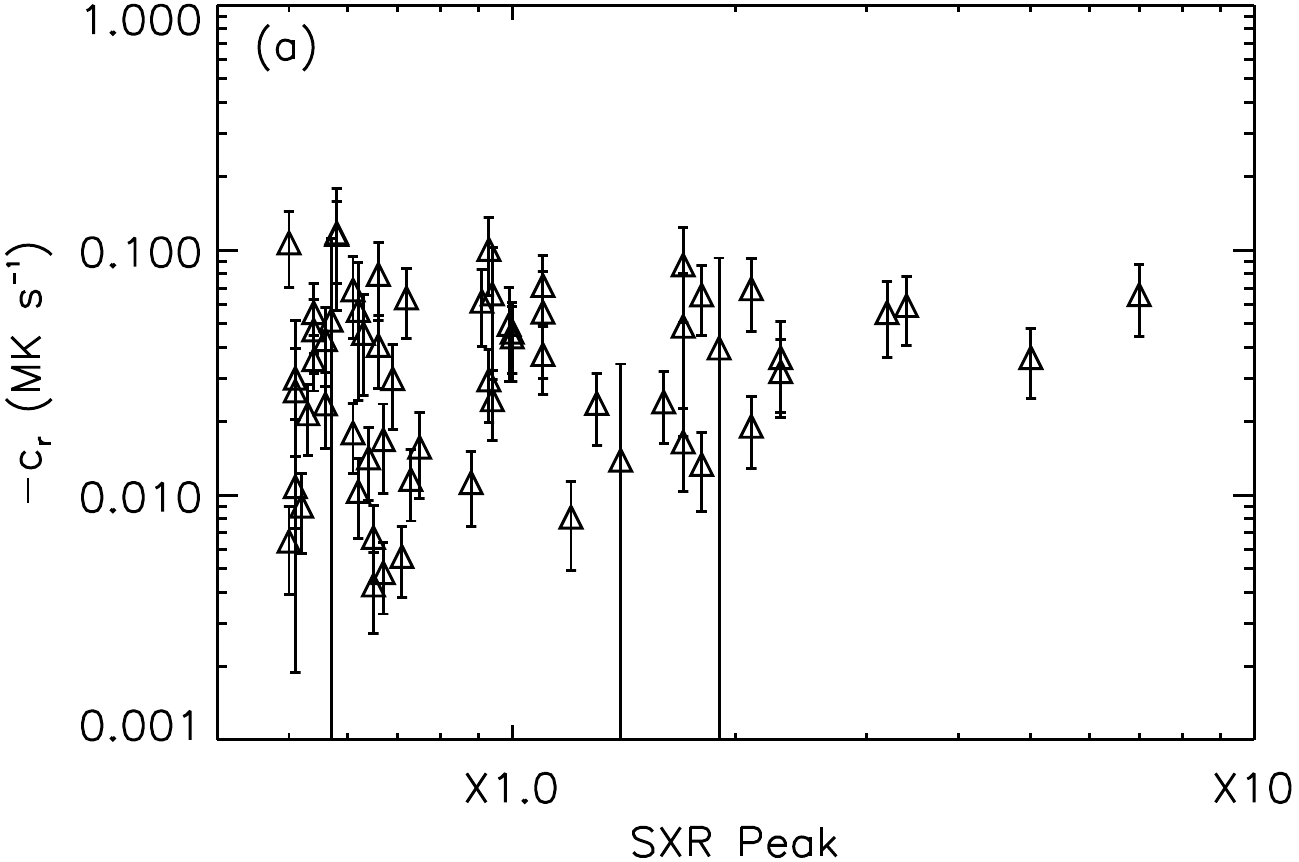}
  \includegraphics[width=0.485\hsize]{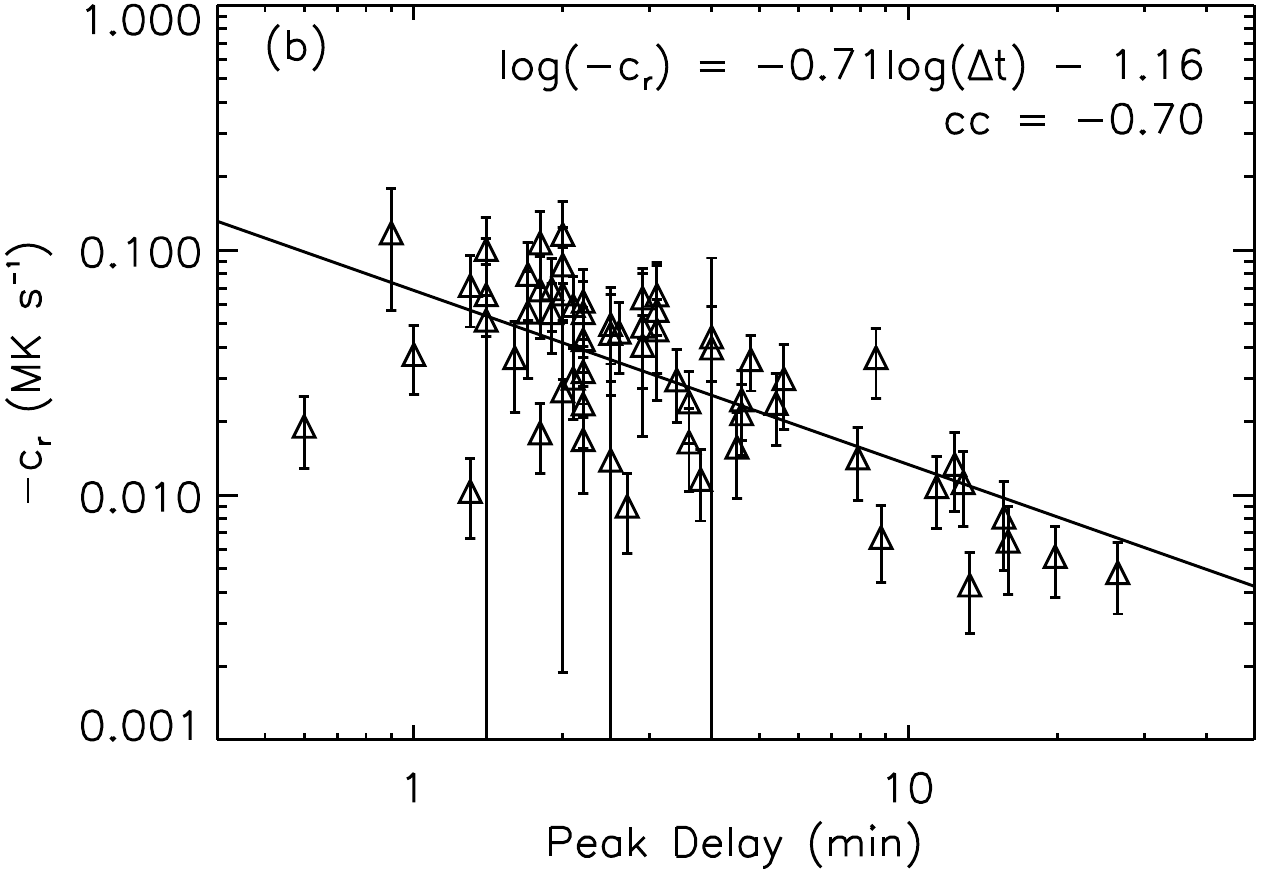}
  \caption{Scatter plots showing (Panel a) the correlation between the linear cooling rate and the SXR peak intensity
  and (Panel b) the correlation between the linear cooling rate and the peak delay time. The solid line in Panel b is a linear fitting to the logarithm values of the parameters.}\label{fg_cr_pd}
\end{figure*}

Not all of the flares have a clear cooling process like the events presented in Sec.~\ref{sec_cases}. 
Those flares (12 events) present alternate cooling and heating signatures in the TDS charts, so that we 
do not try to measure their cooling rates. For the rest, the distribution of the cooling rates has been
displayed in Figure~\ref{fg_cr_pd_dist}b. We find that the cooling rate is about $-0.04$ MK s$^{-1}$ on average, slightly
smaller than the mean value of $-0.035$ MK s$^{-1}$ obtained by \citet{Ryan_etal_2013} for M1.0+ flares. This result
implies that the stronger flares have a faster cooling process. It can be further confirmed in Figure~\ref{fg_cr_pd}a,
in which the trend of the lower limit of the cooling rate is clearly shown. In that plot, one can find that the
cooling rate of M-class flares could be as slow as $-0.004$ MK s$^{-1}$ but that of almost all the X-class flares is faster
than $-0.01$ MK s$^{-1}$.

Furthermore, we compare the cooling rates and the delay of the EUV peaks. A strong power law correlation between them
is found as shown in Figure~\ref{fg_cr_pd}b. The correlation coefficient is about 0.70. The slower cooling rate
does result in a longer delay, confirming the previous speculation that flares experience a coherent cooling process
from SXR emission to EUV emission.

\subsection{Two temperature drift patterns}\label{sec_drift}

\begin{figure*}[tb]
  \centering
  \includegraphics[width=\hsize]{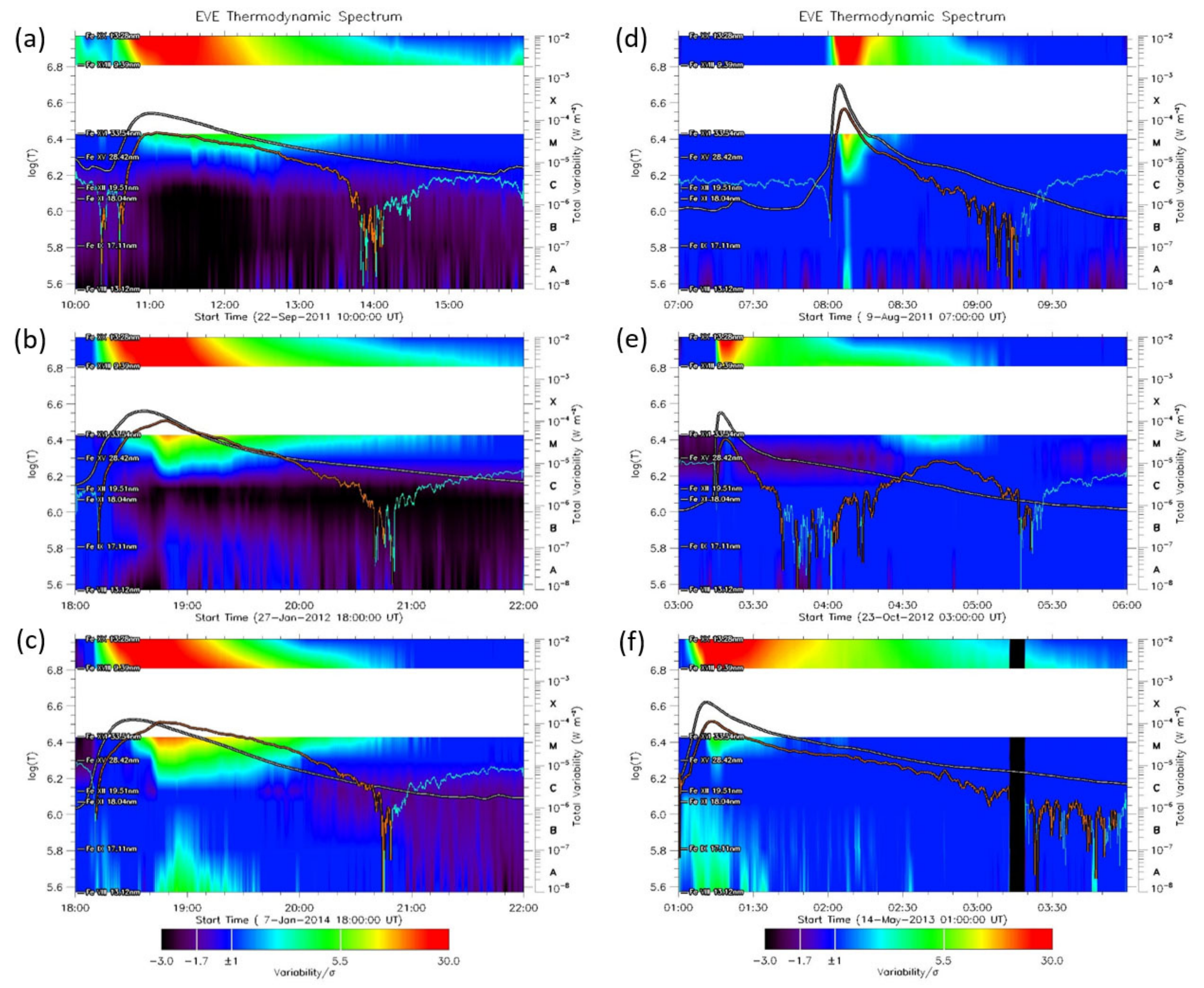}
  \caption{Six example flares showing the Type I drift pattern (left column, Panel a--c) and the Type II drift pattern (right column, Panel d--f). }\label{fg_two_types_examples}
\end{figure*}

By browsing the TDS charts of these M5.0+ flares, we find there are generally
two drift patterns. One pattern shows a drift from higher temperature to lower temperature with time
between the range of $\log(T)$ of about 6.2 and 7.0, looking like a quadrilateral. The other
pattern shows a somewhat different drift; the strongest emissions look like a triangle. In our sample, 52 flares
clearly show such different patterns (as indicated in Table~\ref{tb_m5_flares}).
Figure~\ref{fg_two_types_examples} gives three examples for each pattern.
For convenience, we call the two patterns Type I and II, respectively. The direct cause
of the two types of drift patterns is obvious. For the Type I flares, the enhanced emission from the higher-temperature
line of Fe XX 13.28 nm lasts relatively shorter time than that from the lower-temperature line of Fe
XVIII 9.39 nm, and the situation is reversed for
the Type II flares. It implies that the heating process in Type II flares may be more impulsive so that the plasma can be
heated to higher temperature than that in Type I flares.

\begin{figure*}[tb]
  \centering
  \includegraphics[width=0.485\hsize]{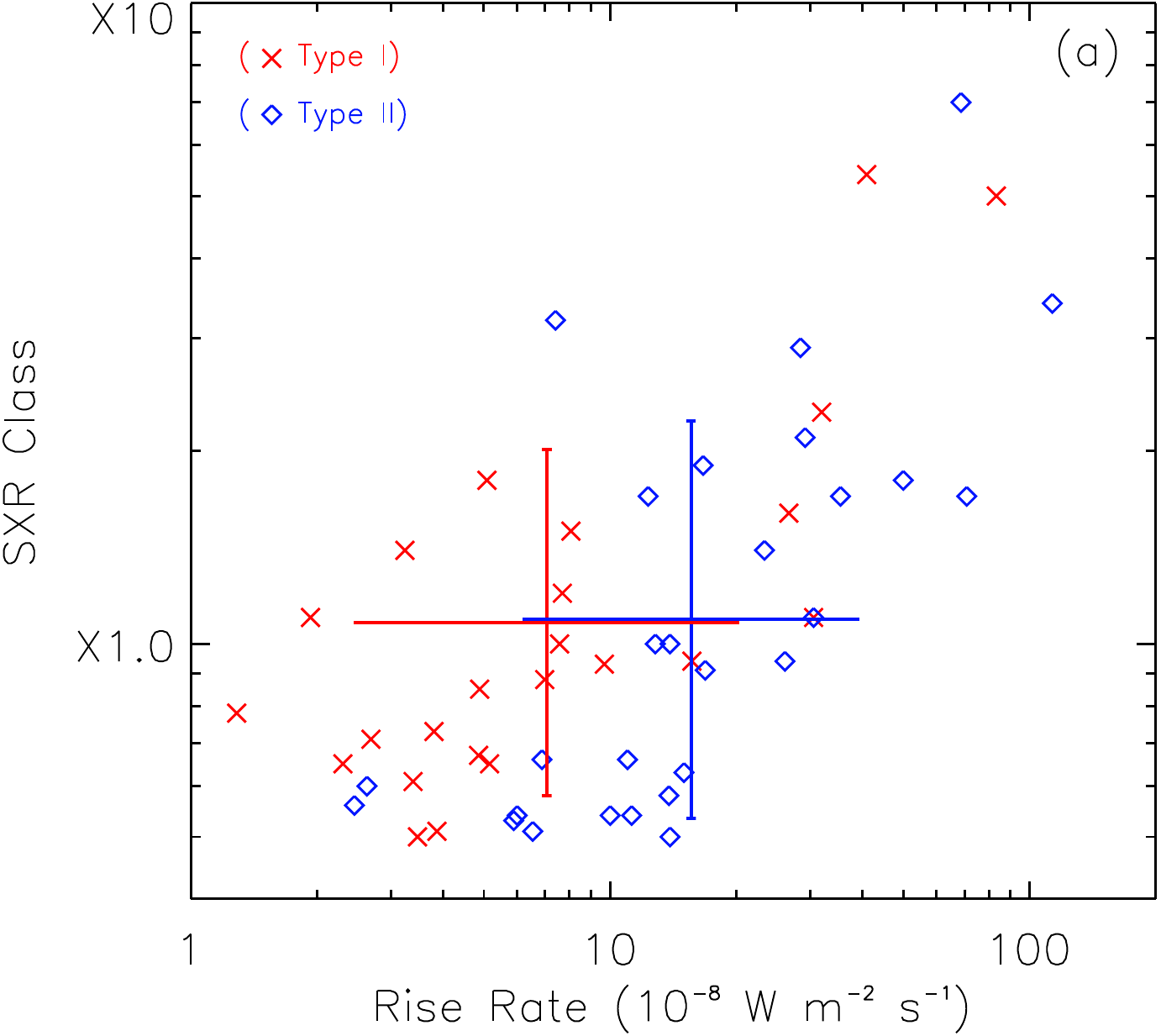}
  \includegraphics[width=0.505\hsize]{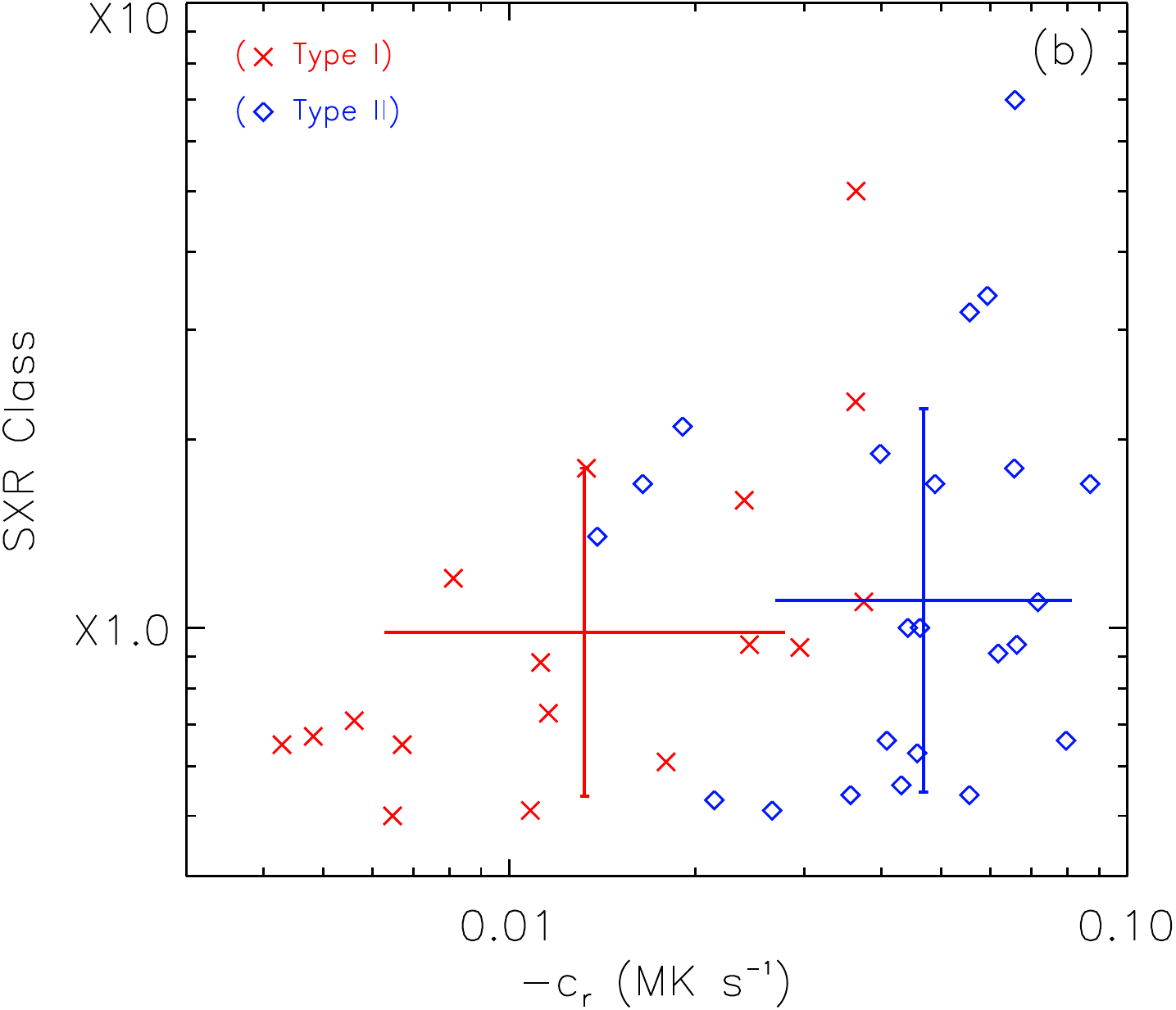}
  \caption{Properties of the flares with different drift patterns (red `x' symbols for Type I and blue diamonds for Type II).
  Panel a shows the scatter plot
  between the SXR rise rate and its peak intensity, and Panel b the scatter plot between the linear cooling
  rate and the SXR peak intensity. The crossed vertical and horizontal lines mark the mean values and the error
bars for the two sets of the data points in the logarithmic scale.}\label{fg_two_types}
\end{figure*}

A statistical analysis is done to the 52 flares, among which 24 flares belong to Type I and 28 flares to Type II.
Here we use the parameter $\frac{\mathrm{Class}}{\mathrm{Peak Time}-\mathrm{Begin Time}}$, which are all inferred from
the GOES SXR emission, as a proxy of the rise rate
of a flare. It is found that, in the plane of the rise rate and SXR peak intensity (Fig.~\ref{fg_two_types}a),
the Type I flares generally locate on the left to the Type II flares, suggesting a slower heating of Type I than Type II flares.
Furthermore, the cooling rates of the two types of the flares are distinct too as shown in Figure~\ref{fg_two_types}b.
Type I flares also tend to have a slower cooling rate or longer cooling time. We think that this is because Type I flares
just heat plasma to a relatively lower temperature in a relatively slower rate than Type II flares, and
therefore for the same amount of released energy, Type I may last longer and show a gradual behavior.

\subsection{The flares with a late phase}
In Section~\ref{sec_cases} we mentioned the eruptiveness of the four flares.
For the two flares without a late phase, the long-duration flare is eruptive, whereas the short-duration
flare is confined. However for the other two flares with the late phase, the eruptiveness is apparently not related to
the duration, as one of them (Case C2) is confined though it lasted for more than 5 hours.
For that particular event, we noticed that it has a strong late-phase, i.e., the peak of
the late phase is higher than that of the main phase. On the contrary, the eruptive flare (Case E2) has a weak late-phase.
We think that the late phase might carry some information of the behavior of a flux rope which is trying to escape
away from the Sun. A strong constraint of the overlying arcades may prevent a flux rope from escaping, and cause
the energy carried by the flux rope to be re-deposited into the thermal emissions which forms a stronger late phase.
Case C2 just fits this scenario as suggested by~\citet{LiuK_etal_2015}. The long-duration confined X-class flare
reported by~\citet{LiuR_etal_2014} also had a significant late phase as shown in their Fig.3. On the other hand,
if the flux rope successfully made its way out, a smaller fraction of its energy will be consumed as thermal emissions, and
a weaker late phase will form.

\begin{figure}[tb]
  \centering
  \includegraphics[width=0.8\hsize]{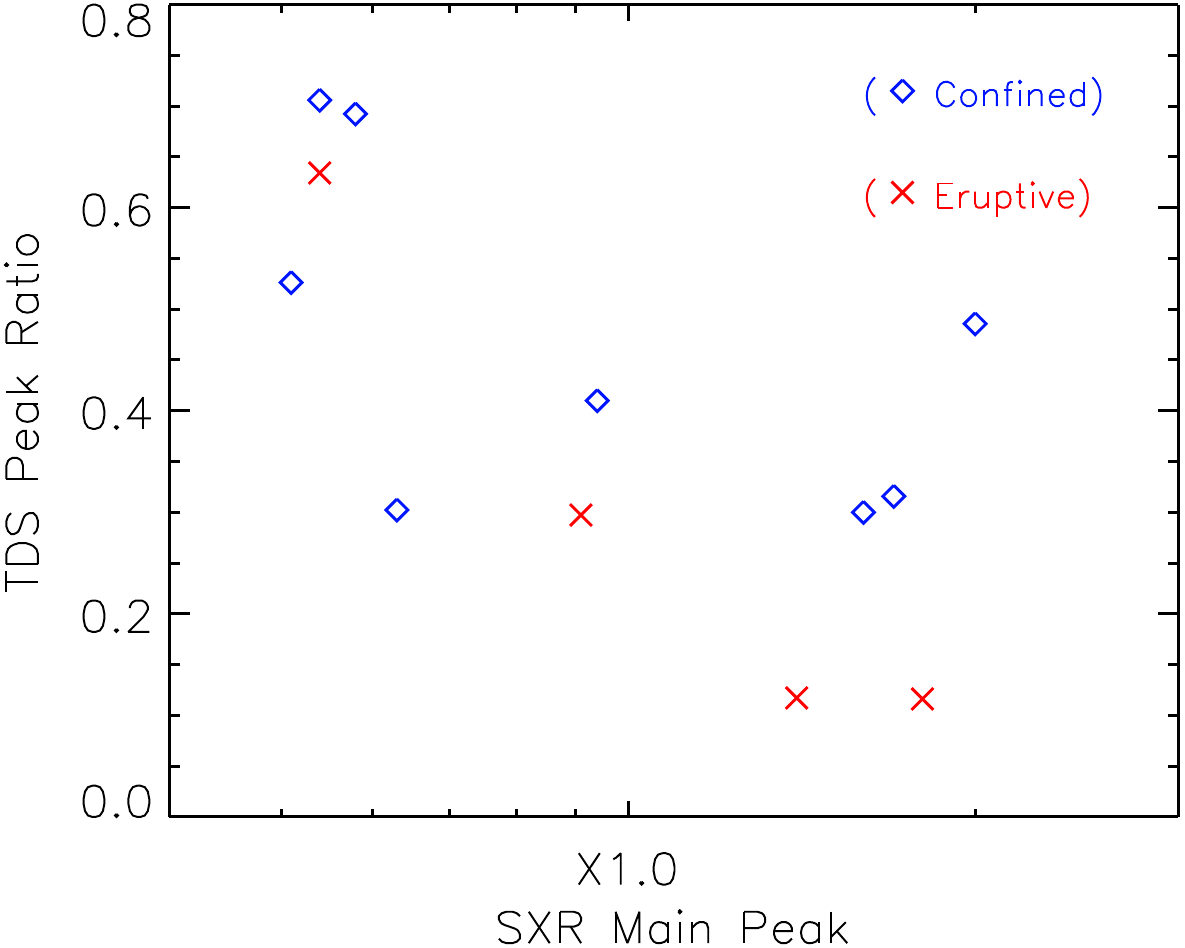}
  \caption{Scatter plot of the 12 late-phase flares. The horizontal axis indicates the flare class
defined by the GOES SXR, and the vertical axis gives the intensity ratio of the first late-phase peak
to the main-phase peak read from the TDS charts. The eruptive flares are marked by red `x' and the
confined flares by blue diamonds.}\label{fg_lp_peaks}
\end{figure}

To check this speculation, we check all the late-phase flares in our sample of M5.0+ flares.
There are 12 flares with a clear late phase, as listed in Table~\ref{tb_latephase}. All these events are confirmed
with SDO/AIA images to ensure that the late-phase peak is related to the main-phase peak. The values of the
late-phase peaks are read from the orange lines in the TDS charts. To make the comparison more convenient, we use the SXR class, e.g.,
`C', `M' and `X', to mark the intensity of the peaks. From Table~\ref{tb_latephase}, one can find that
there could be multiple late-phase peaks (the events L1, L7 and L8)
and the interval between the main-phase peak and the first late-phase peak
could be as short as 33 minutes (the event L6)
or as long as more than 3 hours (the event L12). By comparing the intensity of the
first late-phase peak with that of the main-phase peak read from the TDS charts,
we find a rough correlation between the TDS peak ratio,
which is $\frac{\mathrm{TDS Late Phase Peak 1}}{\mathrm{TDS Main Phase Peak}}$,
and the flare class defined by the SXR main-phase peak, as
shown in Figure~\ref{fg_lp_peaks}. Except the events L8 and L10, the late-phase peaks of the confined flares
are systematically stronger than those of the eruptive flares. Strictly speaking, the pattern in
Figure~\ref{fg_lp_peaks} suggests that a flare with a stronger
late phase must be confined. Although the sample is small, the result does
preliminarily justify our speculation before. An analysis based on a larger sample is worthwhile.

\begin{figure*}[tb]
  \centering
  \includegraphics[width=0.8\hsize]{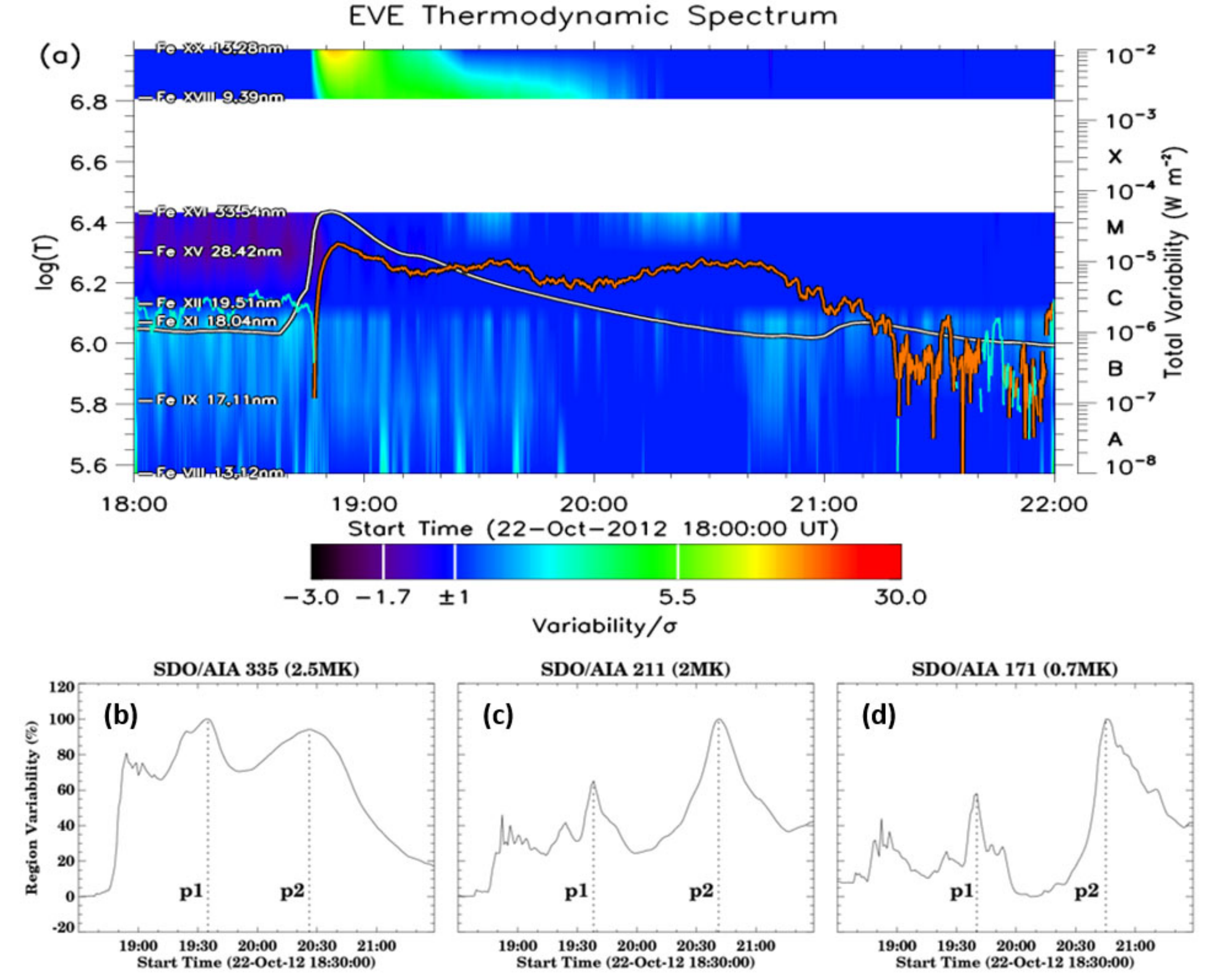}
  \caption{A triple-peak flare occurring on 2012 October 22. (a) TDS chart for the event, in which
the integrated flux indicated by the orange line shows three significant peaks. (b)--(d) The light
curves from the SDO/AIA 33.5 nm, 21.1 nm and 17.1 nm, respectively, which are the integration over
the flaring region as shown in Fig.~\ref{fg_mpeaks}.}\label{fg_mp_tds}
\end{figure*}

\begin{figure*}[tb]
  \centering
  \includegraphics[width=0.8\hsize]{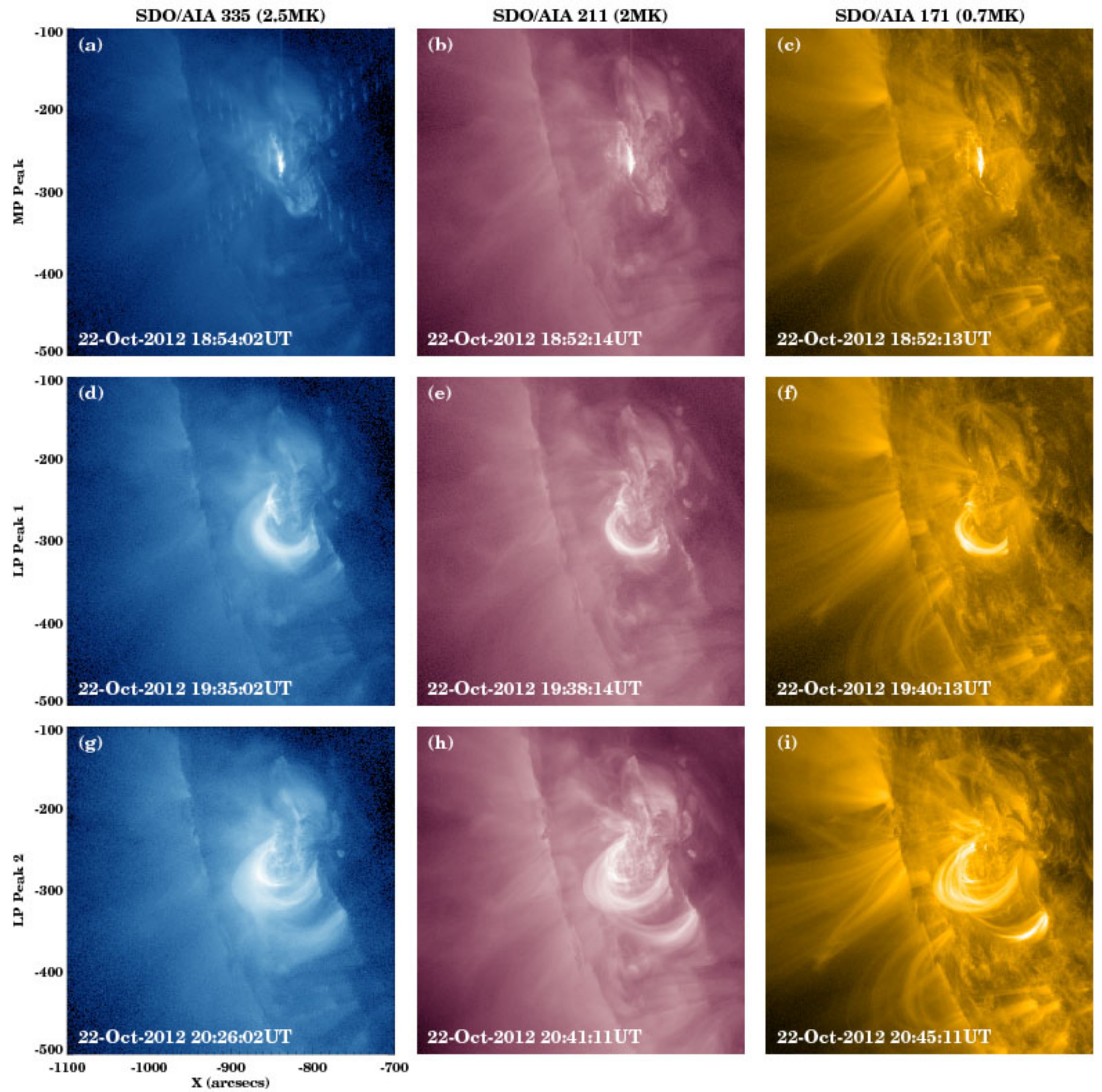}
  \caption{SDO/AIA images showing the flaring region of the 2012 October 22 event. From the left to right,
the panels are for AIA 33.5 nm, 21.1 nm and 17.1 nm, respectively. From the upper to lower, the panels
display the flaring loops/arcades during the main-phase peak, and the two late-phase peaks, respectively.}\label{fg_mpeaks}
\end{figure*}

The multiple late-phase peaks were mentioned before by~\citet{Dai_etal_2013} and \citet{LiuK_etal_2015}.
In our 12 cases, there are three flares with clear multiple peaks during the late phase. The intervals between
these multiple peaks vary from about 55 minutes to about 89 minutes. There is no obvious regulation
among these multiple peaks. A flare with multiple late-phase peaks could be either eruptive or confined.
Figure~\ref{fg_mp_tds} shows a
triple-peak flare on 2012 October 22 (the event L7). In panel (a), one can see one main-phase peak plus two significant
late-phase peaks. Emissions from the line of Fe XVI 33.54 nm most clearly show these peaks. Similar
signatures also could be found in the lines of 21.1 nm and 17.1 nm. Figure~\ref{fg_mpeaks} displays
the flaring region viewed through SDO/AIA 33.5 nm, 21.1 nm and 17.1 nm, respectively, during the
three peaks. The light curves from these three passbands integrated over the flaring region are presented
in the panel (b)--(d) of Figure~\ref{fg_mp_tds}. It confirms that these peaks came from the same event.
Particularly, from Figure~\ref{fg_mpeaks}, the main-phase peak originated from the lowest loops or the core field,
the first late-phase peak from the higher arcades near to the core field, and the second late-phase peak
from the outmost arcades.

\section{Extended TDS charts with the SDO/EVE MEGS-B data}\label{sec_extended}
\subsection{Method}
According to the flare catalog\footnote{\url{http://lasp.colorado.edu/eve/data\_access/evewebdata/interactive/eve\_flare\_catalog.html}}
compiled by the EVE team, MEGS-B captured 82 M-class flares and 6 X-class flares though it did not operate full time.
But not all these flares are completely covered by MEGS-B data in time. That is why we did not use it in the statistical
study of the M5.0+ flares. Here we introduce how we extend the TDS charts with the MEGS-B data, and show some examples.

There are 15 lines from MEGS-B in the EVL product, among which 8 lines,
No.2, 4, 9, 10 and 12--15 in Table~\ref{tb_lines-b}, are significantly blended by other ions based
on our criterion given in Sec.~\ref{sec_line}. All these contaminated lines are discarded. Further, the lines Si XII 49.94 nm
and Ne VIII 77.04 nm have a formation temperature very close to the lines Fe XV 28.42 nm and Fe IX 17.11 nm, respectively,
which have been used in the TDS charts. Thus the two lines are not considered too. In the rest, the lines No.7 and 8,
i.e., O IV 55.44 nm and 79.02 nm, have the same corresponding temperature. After the testing, we find that the two
lines are quite similar. The line O IV 79.02 nm is slightly more sensitive than O IV 55.44 nm (see the deviations listed in
the last two columns of Table~\ref{tb_lines-b}). A larger sensitivity sometimes means that it is easy to get noisy, and
the normalization based on a larger deviation will reduce the significance of real signals. Thus, we choose O IV 55.44 nm
rather than O IV 79.02 nm. Finally, four lines, which
are marked by the asterisks in Table~\ref{tb_lines-b}, are selected to construct the extended TDS charts.

The procedure of generating the extended TDS charts is exactly the same as that in Sec.\ref{sec_data}. One can find the
extended TDS charts at the webpage\\
\url{http://space.ustc.edu.cn/dreams/shm/tds-c09}. Figure~\ref{fg_example_b} shows
an example, which is the same event in Figure~\ref{fg_example_f} so that one can compare them for the difference.
First, it should be noted that the temperature gap between the last two lines in the TDS is large, and we do not try to
interpolate the gap between them. Thus the last line, C III 97.70 nm, is plotted separately as a stripe.

\subsection{Cases}\label{sec_cases-b}
\begin{figure}[tbh]
  \centering
  \includegraphics[width=\hsize]{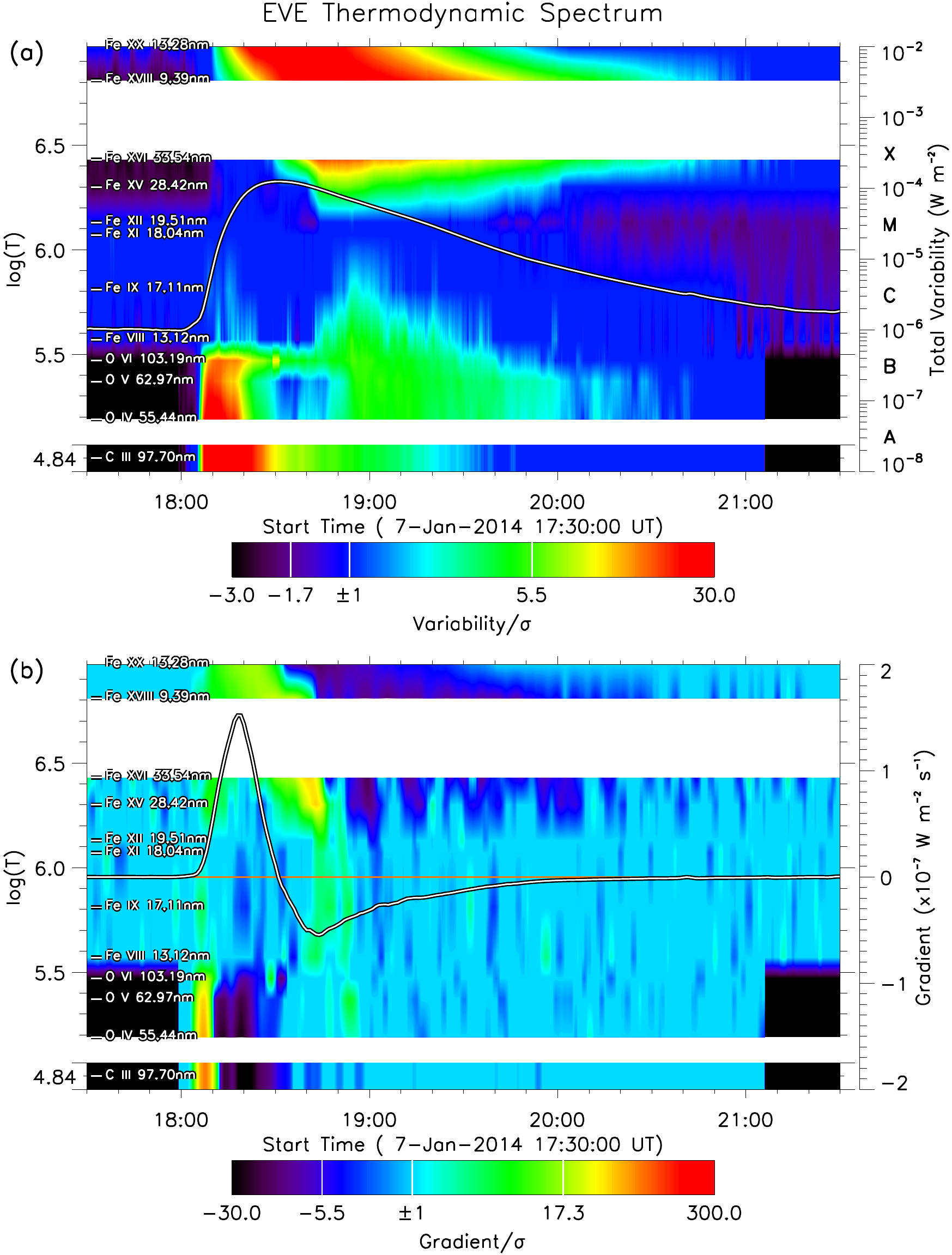}
  \caption{Extended TDS charts of the same event in Fig.\ref{fg_example_f}.}\label{fg_example_b}
\end{figure}

\begin{figure}[tbh]
  \centering
  \includegraphics[width=\hsize]{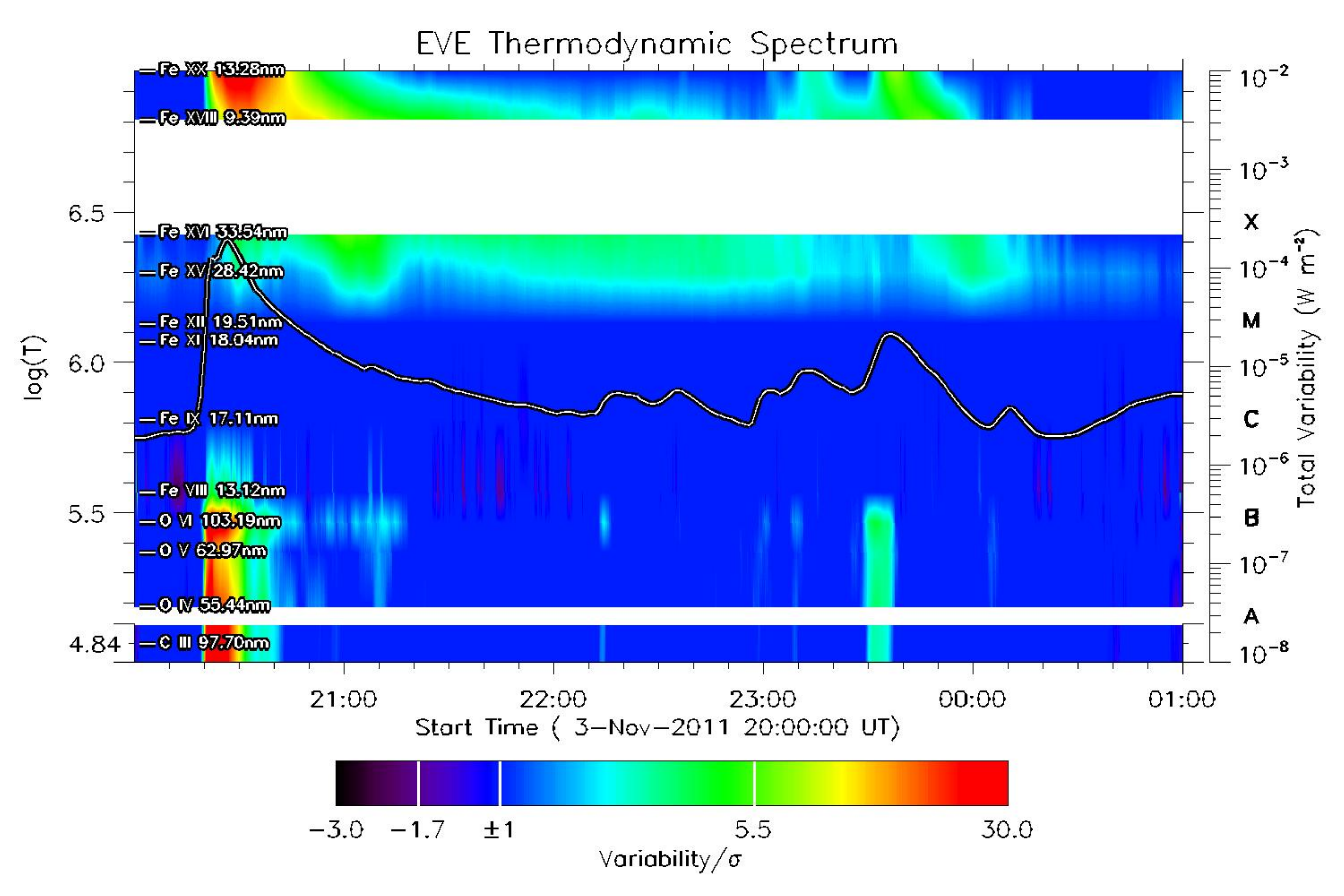}\\
  \includegraphics[width=\hsize]{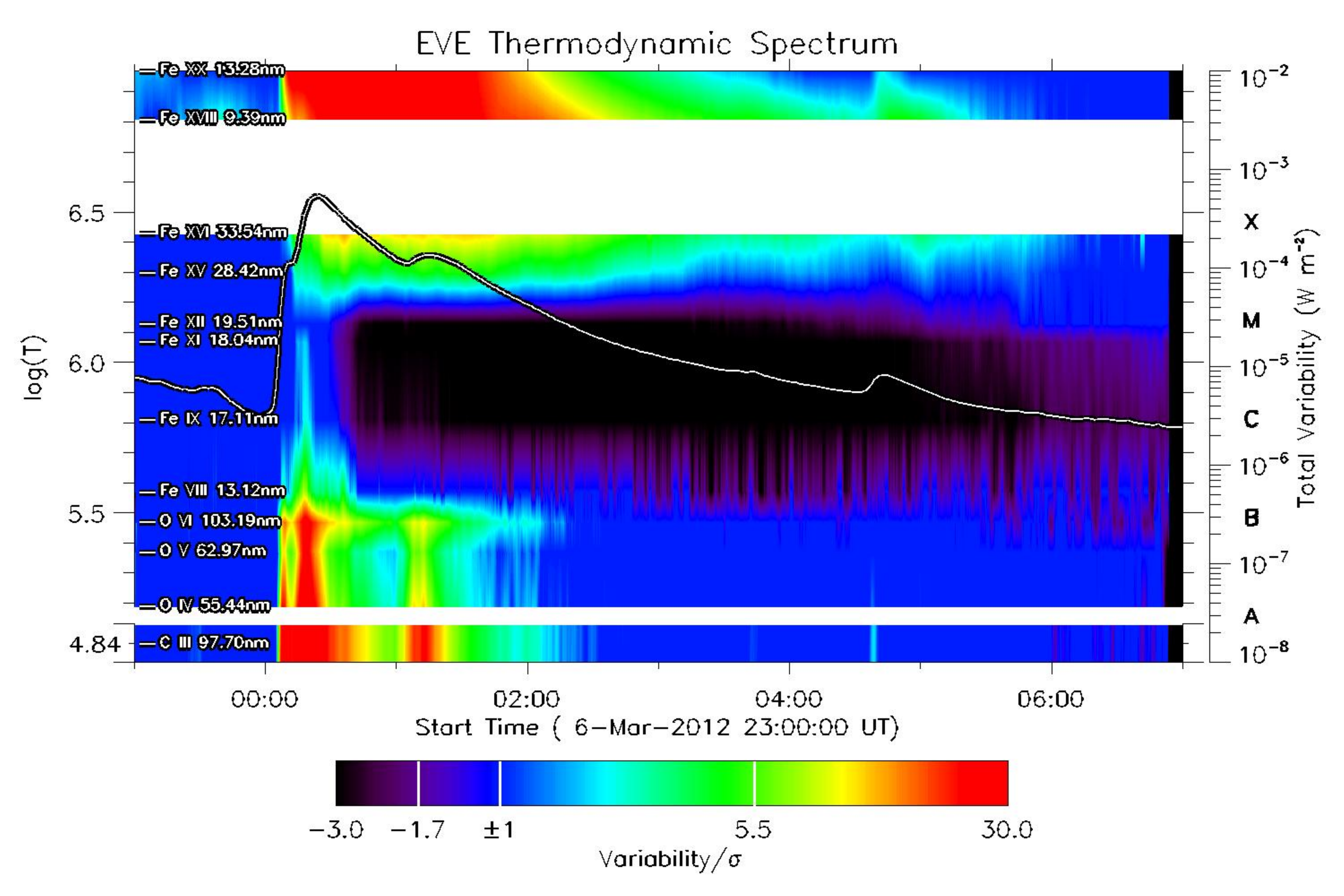}
  \caption{Two X-class flares on 2011 November 3 and 2012 March 7 shown in the extended TDS.}\label{fg_tds-b}
\end{figure}

The extended TDS gives a more complete picture of the thermodynamic process of a solar flare, in which the 
impulsive and gradual phases of a flare~\citep[see the review by][]{Hudson_2011} can be clearly recognized. For the particular event 
on 2014 January 7 (Fig.\ref{fg_example_b}), the enhancement of the emission started first from the temperature below $\log(T)=5.5$, 
which is about 2 minutes ahead of the emission enhancement above the temperature of $\log(T)=6.8$.
The enhancement at the low temperature is stronger but shorter than that at the high temperature.
It is due to the non-thermal heating of the accelerated electrons impacting the 
dense chromosphere and/or the transition region~\citep[see also, e.g.,][]{Milligan_etal_2014}. This is the impulsive phase 
of the flare. Immediately, the impact of non-thermal particles causes the chromospheric evaporation, which transfers the heat to the 
flare loops and heats them up, leading to the thermal phase during which the plasma could be observed more than 
10 MK. This hot plasma then cools down in time as can be seen from the drift pattern in either the variability chart or the gradient
chart. Figure~\ref{fg_tds-b} shows the other two X-class flares which were completely observed by MEGS-B. Similar to the 2014 January 7 event,
the emission enhancement at the low temperature in both events is slightly earlier than that at the high temperature
though the difference in time is not so significant, the strength of the enhancement at the low temperature is stronger
than that at the high temperature, and the duration is relatively shorter.

A noteworthy thing is the variation of the emission between $\log(T)=5.6$ and $6.2$, which is inert to enhancement 
but sensitive to reduction. For the three X-class flares, the emission within that temperature range changed little
at the beginning, but a significant reduction at a later time can be observed for two of them (as seen in Fig.\ref{fg_example_b} and 
the lower panel of Fig.\ref{fg_tds-b}). Both the two flares with the emission reduction
are eruptive, and the other flare is confined. Thus a promising explanation of such a reduction is that the accompanied
CME removes a significant plasma from the corona which mainly locates within the temperature range from about $\log(T)=5.6$ to $6.2$. 
A statistical survey on the CME signatures in TDS charts will be done in a separated paper. 
So far, it is still a mystery why the emission corresponding to that temperature range is hard to be enhanced during a flare.
Whether it is simply a weakness of our TDS visualization method or there is some unknown physical mechanism is worthy of
further study.

\section{Summary and discussion}\label{sec_summary}

In this work, we present a new method to show the thermodynamic processes of solar activities,
the so-called thermodynamic spectrum chart, which is constructed based on the SDO/EVE data. 
The TDS charts provide a global view of the thermal processes during solar flares, especially when 
both MEGS-A and MEGS-B data are incorporated. By investigating four flares of different types, 
we present in details how to read the flare information from the TDS charts.
Reading from the charts, we are able to easily recognize if there is a late phase following 
a main phase of a flare, and able to learn the start, peak and end times of the flare as well 
as the drift of the temperature of the heated plasma during the flare. The advantages of TDS may not
only be in studies of flares but also CMEs, and there are still some unclear signatures 
in TDS as discussed in Sec.\ref{sec_cases-b}.

We apply the TDS method to all the M5.0+ flares during the EVE/MEGS-A's 5-year lifetime.
First, we measure the delay time of the EUV peaks and the cooling rates of these flares. It is found that
EUV peaks are always behind the SXR peaks, and the mean value of the delay time is about 5 minutes. The
stronger flares tend to have a faster cooling rate, and the mean value of the cooling rate is about $-0.04$ MK s$^{-1}$.
There is a clear power law correlation between the cooling rates
and the peak delay times, which suggests a coherent cooling process of flares within the temperature range from
SXR down to EUV emissions.

Second, we find that there are two temperature drift patterns of flares in the TDS charts, called Type I and Type II.
Type I flares show a quadrilateral-like drift mode from high to low temperature with time, and the others shows a triangle-like
drift mode. The statistical analysis reveals that Type I flares are generally more gradual and their heating processes
are more durable than Type II flares, whereas Type II flares are impulsive and more plasma at higher temperature may be
heated.

Third, the strength of the late-phase peak is relevant to the eruptiveness of a flare. A rough
correlation could be found between the TDS peak ratio and the SXR flare class, suggesting that
a strong late-phase is probably caused by a confined flare, during which the energy carried by the
flux rope, that was trying to erupt out, is re-deposited into the thermal emissions. This result gives us new clues
to understand the energy partition and transfer process during the attempt of a flux rope eruption.

\citet{Warren_etal_2013} constructed similar charts by
computing differential emission measure (DEM) distribution. Their DEM method gave the TDS
within the temperature range of $\log(T)=6.3$ to $7.8$ without gaps. As mentioned in their
paper, the weakness of such charts is that the uncertainties in DEM are hard to assess, which is because it is
model-based and many assumptions have to be made. Compared with the DEM charts, our charts are
almost model-free, and the temperature range is from $\log(T)=5.57$ (or 4.84 if MEGS-B data are available) 
to $6.97$. Low temperature resolution
might be the weakness of our current charts, but it could be improved by incorporating more emission lines.
Besides, the temperature indicated in our TDS stands for the peak formation temperature of a line, which may
deviate away from real temperature of the emitting plasma particularly in non-isothermal circumstance.
At this point, one should be caution to use TDS to interpret the thermal processes. However,
the comparison between the TDS and DEM charts of the 5 events shown in \citet{Warren_etal_2013} paper
(Fig.8 and 9 there) suggests that the deviation may not be significant as their patterns look
similar in the common temperature range (One can check our online website to make the comparison).
The two kinds of charts could be usefully complementary to each other.

The technique of the TDS presented here may not be only limited to solar physics.
As mentioned in Introduction, since the first rocket-based experiments in 1960s, there have been lots of EUV
observations of distant astronomical objects, e.g., stars, in the universe. Due to the far distance,
there is no detailed imaging data of those stars. Thus, the TDS
technique established here provides a potential approach to learn the eruptive activities over there.

\acknowledgments{We acknowledge the use of the data from SDO, STEREO, SOHO
and GOES spacecraft. SDO is a mission of NASA's Living With a Star Program, STEREO is the
third mission in NASA's Solar Terrestrial Probes programme, and SOHO is a mission of
international cooperation between ESA and NASA. The TDS charts for all the events involved 
in this study could be found at \url{http://space.ustc.edu.cn/dreams/shm/tds} (the MEGS-A-only TDS) 
and \url{http://space.ustc.edu.cn/dreams/shm/tds-c09} (the extended TDS).
This work is supported by grants from NSFC (41131065, 41574165, 41421063, 41274173, 41222031, 41404134 and 41474151),
CAS (Key Research Program KZZD-EW-01 and 100-Talent Program), MOST 973 key project (2011CB811403) and 
the fundamental research funds for the central universities.}

\bibliographystyle{agu}
\bibliography{../../ahareference}

\clearpage

\begin{table*}
\begin{center}
\footnotesize
\caption{Information of emission lines provided by MEGS-A$^*$}\label{tb_lines}
\begin{tabular}{c|cccccccc}
\hline
No. & Ions & $\lambda_{min}$ & $\lambda_{max}$ & $\lambda_{cen}$ & $\log{(T)}$ & $\sigma_{I\pm}$ & $\sigma_{G\pm}$ \\
& & nm & nm & nm & $\log$(K$^\circ$) & $\times10^{-7}$ W m$^{-2}$ & $\times10^{-9}$ W m$^{-2}$ s$^{-1}$ \\
\hline
1  & Fe XX$^*$    & 13.23 & 13.32 & 13.29  &       6.97  & $+9.3/-0.2$ & $+2.3/-1.6$\\
2  & Fe XVIII$^*$ & 9.33  & 9.43  & 9.39   &       6.81  & $+5.5/-1.7$ & $+0.6/-0.4$\\
3  & Fe XVI$^*$   & 33.47 & 33.61 &  33.54 &       6.43  & $+28.2/-21.8$ & $+1.9/-1.7$\\
4  & Fe XVI   & 36.02 & 36.20 &  36.08 &       6.43  & &\\
5  & Fe XV$^*$    & 28.30 & 28.50 &  28.42 &       6.30  & $+30.1/-26.8$ & $+1.3/-1.2$ \\
6  & Fe XIV   & 21.07 & 21.20 &  21.13 &       6.27  & &\\
7  & Fe XIII  & 20.14 & 20.32 &  20.20 &       6.19  & & \\
8  & Fe XII$^*$   & 19.43 & 19.61 &  19.51 &       6.13  & $+17.9/-10.1$ & $+1.8/-1.7$ \\
9  & Fe XI$^*$    & 17.96 & 18.15 &  18.04 &       6.07  & $+11.2/-12.2$ & $+1.2/-1.1$ \\
10 & Fe X     & 17.63 & 17.83 &  17.72 &       5.99  & &\\
11 & Mg IX    & 36.71 & 36.89 &  36.81 &       5.99  & &\\
12 & Fe IX$^*$    & 17.02 & 17.24 &  17.11 &       5.81  & $+14.4/-14.5$ & $+1.5/-1.5$ \\
13 & Fe VIII$^*$  & 13.04 & 13.17 &  13.12 &       5.57  & $+0.5/-0.5$ & $+0.1/-0.1$ \\
14 & He II    & 25.55 & 25.68 &  25.63 &       4.75  & &\\
15 & He II    & 30.25 & 30.50 &  30.38 &       4.70  & &\\
\hline
\end{tabular}\\
$^*$ Column 3 to 5 give the wavelength range and peak wavelength of each spectral line, Column 6 lists the
corresponding formation temperature, and the last two columns give the deviations of the variabilities and
gradients of final selected spectral lines (see Sec.\ref{sec_data} for more details).
\end{center}
\end{table*}

\begin{table*}
\begin{center}
\footnotesize
\caption{Information of the four flares of interest$^*$}\label{tb_flares}
\begin{tabular}{c|ccccc|cccc|l}
\hline
No. & \multicolumn{5}{c|}{GOES SXR} & \multicolumn{4}{c|}{EVE TDS} & Ref. \\
\cline{2-6}\cline{7-10}
& Date & Begin & Peak & Dur. & Class & Begin & Peak & LP & Dur. & \\
& & UT & UT & Min. & & UT & UT & UT & Min. & \\
\hline
C1  & 2011-09-08 & 15:32 & 15:46 & 20 & M6.7 & 15:32 & 15:48 & No & $\sim40$ & \\
C2  & 2010-11-05 & 12:43 & 13:29 & 83 & M1.0 & 13:15 & 13:32 & 16:42 & $>300$ & W11, C12, L15 \\
E1  & 2011-03-08 & 03:37 & 03:58 & 48 & M1.5 & 13:40 & 14:17 & No & $\sim100$ & Z12, R13 \\
E2  & 2010-10-16 & 19:07 & 19:12 & 5  & M3.0 & 19:09 & 19:13 & 20:26 & $>110$ & W11, L13, R13 \\
\hline
\end{tabular}\\
$^*$ The first column indicates confined or eruptive. The next 5 columns list the flare parameters
based on GOES SXR reports, and the following four columns based on EVE TDS charts.
In the column 5 and 10, `Dur.'
means duration in units of minutes. Column 9, `LP', lists the peak time of the late phase if any.
The last column lists some references, in which the flares were investigated. W11 refers to \citet{Woods_etal_2011},
C12 to \citet{Chamberlin_etal_2012}, Z12 to \citet{Zhang_etal_2012},
L13 to \citet{LiuK_etal_2013}, R13 to \citet{Ryan_etal_2013}, and L15 to \citet{LiuK_etal_2015}.
\end{center}
\end{table*}

\clearpage

\begin{table*}
\tiny
\setlength{\tabcolsep}{3pt}
\begin{center}
\caption{X-ray flares equal to or greater than M5.0 from 2010 May to 2014 May$^*$}\label{tb_m5_flares}
\begin{tabular}{c| c | c  c | c  c c | c | c ||c| c | c  c | c  c c | c |c }
\hline
No. & Date & \multicolumn{2}{|c }{SXR} & \multicolumn{3}{|c| }{TDS} & Peak & Type & No. & Date & \multicolumn{2}{|c }{SXR} & \multicolumn{3}{|c| }{TDS} & Peak & Type \\
\cline{3-7}\cline{12-16}
 &  & Time & Class & Time & Class & $c_r$ & delay & &  &  & Time & Class & Time & Class & $c_r$ & delay & \\
 &  & (UT) & & (UT) & & (MK s$^{-1}$) & (min) & &  &  & (UT) & & (UT) & & (MK s$^{-1}$) & (min) & \\
\hline
 1& 2010-11-06 & 15:36 & M5.4 & 15:41 & M4.1 & -0.036 &  4.8 &II  & 38& 2012-10-20 & 18:14 & M9.1 & 18:16 & M3.7 & -0.062 &  2.2 &II \\
 2& 2011-02-13 & 17:38 & M6.6 & 17:41 & M4.3 & -0.041 &  2.9 &II  & 39& 2012-10-22 & 18:51 & M5.1 & 18:53 & M1.8 & -0.027 &  2.0 &II \\
 3& 2011-02-15 & 01:56 & X2.3 & 01:58 & M9.0 & -0.032 &  2.2 &    & 40& 2012-10-23 & 03:17 & X1.7 & 03:19 & M3.8 & -0.087 &  2.0 &II \\
 4& 2011-02-18 & 10:11 & M6.6 & 10:13 & M3.5 & -0.080 &  1.7 &II  & 41& 2012-11-13 & 02:04 & M6.1 & 02:06 & M3.6 & -0.069 &  1.8 &   \\
 5& 2011-03-08 & 10:44 & M5.4 & 10:47 & M1.7 & -0.047 &  3.1 &    & 42& 2013-04-11 & 07:16 & M6.5 & 07:25 & M5.8 & -0.007 &  8.8 &I  \\
 6& 2011-03-09 & 23:23 & X1.6 & 23:27 & M8.0 & -0.024 &  3.6 &I   & 43& 2013-05-03 & 17:32 & M5.7 & 17:33 & M3.9 & -0.052 &  1.4 &   \\
 7& 2011-07-30 & 02:09 & M9.3 & 02:10 & M6.2 & -0.101 &  1.4 &    & 44& 2013-05-13 & 02:16 & X1.7 & 02:20 & M5.8 & -0.016 &  3.6 &II \\
 8& 2011-08-03 & 13:47 & M6.1 & 13:49 & M5.0 & -0.018 &  1.8 &I   & 45& 2013-05-13 & 16:05 & X2.9 & 16:07 & M6.5 &        &  1.6 &II \\
 9& 2011-08-04 & 03:57 & M9.3 & 04:00 & M7.4 & -0.030 &  3.4 &I   & 46& 2013-05-14 & 01:11 & X3.2 & 01:13 & X1.1 & -0.056 &  2.2 &II \\
10& 2011-08-09 & 08:05 & X7.0 & 08:06 & X1.9 & -0.066 &  1.4 &II  & 47& 2013-05-15 & 01:48 & X1.3 & 01:53 & M7.1 & -0.024 &  5.4 &   \\
11& 2011-09-06 & 01:50 & M5.4 & 01:51 & M2.8 &        &  1.1 &II  & 48& 2013-05-22 & 13:32 & M5.0 & 13:48 & M5.1 & -0.006 & 15.9 &I  \\
12& 2011-09-06 & 22:20 & X2.1 & 22:22 & M9.6 & -0.069 &  1.9 &    & 49& 2013-06-07 & 22:49 & M6.0 & 22:49 & M3.0 &        &  0.5 &II \\
13& 2011-09-07 & 22:38 & X1.8 & 22:41 & M8.0 & -0.066 &  3.1 &II  & 50& 2013-10-24 & 00:30 & M9.4 & 00:35 & M7.3 & -0.024 &  4.6 &I  \\
14& 2011-09-08 & 15:46 & M6.7 & 15:48 & M2.4 & -0.017 &  2.2 &    & 51& 2013-10-25 & 08:01 & X1.7 & 08:04 & M4.5 & -0.049 &  2.9 &II \\
15& 2011-09-22 & 11:00 & X1.5 & 11:05 & M5.2 &        &  5.1 &I   & 52& 2013-10-25 & 15:03 & X2.1 & 15:04 & M5.9 & -0.019 &  0.6 &II \\
16& 2011-09-24 & 09:40 & X1.9 & 09:44 & M5.5 & -0.040 &  4.0 &II  & 53& 2013-10-28 & 02:03 & X1.0 & 02:06 & M4.4 &        &  2.8 &I  \\
17& 2011-09-24 & 13:17 & M7.1 & 13:37 & M5.4 & -0.006 & 19.8 &I   & 54& 2013-10-28 & 04:41 & M5.1 & 04:43 & M1.6 & -0.030 &  2.1 &   \\
18& 2011-09-24 & 20:36 & M5.8 & 20:38 & M2.6 & -0.116 &  2.0 &    & 55& 2013-10-29 & 21:54 & X2.3 & 21:56 & M5.9 & -0.036 &  1.6 &I  \\
19& 2011-09-25 & 04:50 & M7.5 & 04:54 & M5.2 & -0.016 &  4.5 &    & 56& 2013-11-01 & 19:53 & M6.3 & 19:55 & M4.3 & -0.046 &  2.5 &II \\
20& 2011-11-03 & 20:27 & X2.0 & 20:29 & M7.0 &        &  2.3 &    & 57& 2013-11-03 & 05:22 & M5.0 & 05:24 & M3.2 & -0.107 &  1.8 &II \\
21& 2012-01-23 & 03:59 & M8.8 & 04:12 & M8.1 & -0.011 & 12.9 &I   & 58& 2013-11-05 & 22:12 & X3.4 & 22:14 & M7.7 & -0.059 &  2.1 &II \\
22& 2012-01-27 & 18:36 & X1.8 & 18:48 & X1.1 & -0.013 & 12.4 &I   & 59& 2013-11-08 & 04:26 & X1.1 & 04:27 & M4.9 & -0.072 &  1.3 &II \\
23& 2012-03-05 & 04:05 & X1.1 & 04:16 & M6.8 &        & 10.5 &I   & 60& 2013-11-10 & 05:14 & X1.1 & 05:15 & M6.4 & -0.037 &  1.0 &I  \\
24& 2012-03-07 & 00:24 & X5.4 & 00:27 & X1.6 &        &  3.1 &I   & 61& 2013-11-19 & 10:26 & X1.0 & 10:30 & M2.9 & -0.044 &  4.0 &II \\
25& 2012-03-09 & 03:53 & M6.4 & 04:01 & M6.1 & -0.014 &  7.9 &    & 62& 2013-12-31 & 21:58 & M6.5 & 22:00 & M4.7 &        &  2.2 &   \\
26& 2012-03-10 & 17:44 & M8.5 & 17:53 & M5.3 &        &  9.1 &I   & 63& 2014-01-01 & 18:52 & M9.9 & 18:55 & M6.9 & -0.050 &  2.5 &   \\
27& 2012-05-10 & 04:18 & M5.8 & 04:19 & M3.0 & -0.117 &  0.9 &II  & 64& 2014-01-07 & 10:13 & M7.2 & 10:16 & M3.5 & -0.064 &  2.9 &   \\
28& 2012-05-17 & 01:47 & M5.1 & 01:58 & M4.2 & -0.011 & 11.4 &I   & 65& 2014-01-07 & 18:30 & X1.2 & 18:46 & X1.1 & -0.008 & 15.6 &I  \\
29& 2012-07-02 & 10:52 & M5.6 & 10:54 & M5.2 & -0.024 &  2.2 &    & 66& 2014-01-30 & 16:11 & M6.7 & 16:37 & M5.3 & -0.005 & 26.5 &I  \\
30& 2012-07-04 & 09:55 & M5.4 & 09:57 & M3.5 & -0.056 &  1.9 &II  & 67& 2014-02-04 & 04:00 & M5.2 & 04:03 & M3.1 & -0.009 &  2.7 &   \\
31& 2012-07-05 & 11:44 & M6.2 & 11:47 & M2.5 & -0.057 &  3.1 &    & 68& 2014-02-25 & 00:49 & X5.0 & 00:58 & X1.2 & -0.036 &  8.6 &I  \\
32& 2012-07-06 & 23:08 & X1.1 & 23:10 & M4.7 & -0.056 &  1.7 &    & 69& 2014-03-12 & 22:34 & M9.4 & 22:36 & M2.0 & -0.066 &  2.0 &II \\
33& 2012-07-08 & 16:32 & M6.9 & 16:38 & M2.9 & -0.030 &  5.6 &    & 70& 2014-03-29 & 17:48 & X1.0 & 17:51 & M5.4 & -0.046 &  2.6 &II \\
34& 2012-07-12 & 16:49 & X1.4 & 17:08 & M6.9 &        & 18.8 &I   & 71& 2014-04-02 & 14:05 & M6.5 & 14:18 & M4.3 & -0.004 & 13.3 &I  \\
35& 2012-07-19 & 05:58 & M7.8 & 05:57 & M2.0 &        & -0.9 &I   & 72& 2014-04-18 & 13:03 & M7.3 & 13:07 & M6.6 & -0.012 &  3.8 &I  \\
36& 2012-07-28 & 20:56 & M6.2 & 20:57 & M5.0 & -0.010 &  1.3 &    & 73& 2014-04-25 & 00:27 & X1.4 & 00:30 & M2.9 & -0.014 &  2.5 &II \\
37& 2012-08-18 & 01:02 & M5.6 & 01:04 & M1.7 & -0.043 &  2.2 &II  & 74& 2014-05-08 & 10:07 & M5.3 & 10:12 & M2.3 & -0.021 &  4.6 &II \\
\hline
\end{tabular}\\
$^*$ In this table, $c_r$ is the linear cooling rate measured from TDS charts through the method introduced in Sec.\ref{sec_cases}.
For those flares that cannot be simply measured by a linear cooling process, we leave them blank. The column of `Peak delay'
means the delay time of the EUV peaks with respect to the associated SXR peaks. The column of `Type' indicates if the flare shows a
clear Type I or II drift in its EVE TDS chart (refer to Sec.\ref{sec_drift} for more details).
\end{center}
\end{table*}

\begin{table*}
\footnotesize
\setlength{\tabcolsep}{3pt}
\begin{center}
\caption{Flares with a clear late phase}\label{tb_latephase}
\begin{tabular}{c| c | c  c | c  c | c  c | c  c | c  c | c }
\hline
No. & Date & \multicolumn{2}{|c }{SXR main phase} & \multicolumn{2}{|c }{TDS main phase} & \multicolumn{6}{|c| }{TDS late phase} & Eruptive$^*$ \\
\cline{3-12}
 &  & Time & Class & Time & Class & Time 1 & Peak 1 & Time 2 & Peak 2 & Time 3 & Peak 3 & \\
\hline

L1& 2010-11-06 & 15:36 & M5.4 & 15:41 & M4.1 & 16:38 & M2.6 & 18:09 & M2.6 & 19:28 & M2.4 & Y \\
L2& 2011-03-08 & 10:44 & M5.4 & 10:47 & M1.7 & 12:16 & M1.2 & - & - & - & - & N \\
L3& 2011-03-09 & 23:23 & X1.6 & 23:27 & M8.0 & 00:01 & M2.4 & - & - & - & - & N \\
L4& 2011-09-07 & 22:38 & X1.8 & 22:41 & M8.0 & 23:52 & C9.3 & - & - & - & - & Y \\
L5& 2011-09-24 & 20:36 & M5.8 & 20:38 & M2.6 & 21:24 & M1.8 & - & - & - & - & N \\
L6& 2011-11-03 & 20:27 & X2.0 & 20:29 & M7.0 & 21:02 & M3.4 & - & - & - & - & N \\
L7& 2012-10-20 & 18:14 & M9.1 & 18:16 & M3.7 & 19:48 & M1.1 & 21:16 & M1.4 & - & - & Y \\
L8& 2012-10-22 & 18:51 & M5.1 & 18:53 & M1.8 & 19:33 & M1.0 & 20:28 & M1.1 & - & - & N \\
L9& 2012-10-23 & 03:17 & X1.7 & 03:19 & M3.8 & 04:46 & M1.2 & - & - & - & - & N \\
L10& 2013-11-01 & 19:53 & M6.3 & 19:55 & M4.3 & 21:51 & M1.3 & - & - & - & - & N \\
L11& 2014-03-12 & 22:34 & M9.4 & 22:36 & M2.0 & 00:33 & C8.2 & - & - & - & - & N \\
L12& 2014-04-25 & 00:27 & X1.4 & 00:30 & M2.9 & 03:42 & C3.4 & - & - & - & - & Y \\
\hline
\end{tabular}\\
$^*$ `Y' means that the flare is associated with a CME, and `N' means there is no CME.
\end{center}
\end{table*}

\begin{table*}
\begin{center}
\footnotesize
\caption{Information of emission lines provided by MEGS-B$^*$}\label{tb_lines-b}
\begin{tabular}{c|cccccccc}
\hline
No. & Ions & $\lambda_{min}$ & $\lambda_{max}$ & $\lambda_{cen}$ & $\log{(T)}$ & $\sigma_{I\pm}$ & $\sigma_{G\pm}$ \\
& & nm & nm & nm & $\log$(K$^\circ$) & $\times10^{-7}$ W m$^{-2}$ & $\times10^{-9}$ W m$^{-2}$ s$^{-1}$ \\
\hline
1  & Si XII  &   49.84 &  50.04 &  49.94 & 6.29  & & \\
2  & Mg X    &   62.28 &  62.68 &  62.49 & 6.05  & & \\
3  & Ne VIII &   76.90 &  77.18 &  77.04 & 5.81  & & \\
4  & Ne VII  &   46.32 &  46.74 &  46.52 & 5.71  & &\\
5  & O VI$^*$&  103.10 & 103.32 & 103.19 & 5.47  & $+19.2/-11.7$ & $+5.5/-4.2$ \\
6  & O V$^*$ &   62.74 &  63.18 &  62.97 & 5.37  & $+8.4/-5.7$ & $+1.7/-1.3$\\
7  & O IV$^*$&   55.20 &  55.64 &  55.44 & 5.19  & $+2.8/-1.7$ & $+0.8/-0.6$ \\
8  & O IV    &   78.90 &  79.14 &  79.02 & 5.19  & $+3.1/-1.9$ & $+1.3/-1.2$ \\
9  & O III   &   52.42 &  52.72 &  52.58 & 4.92  & &  \\
10 & O III   &   59.84 &  60.14 &  59.96 & 4.92  & &\\
11 & C III$^*$&   97.56 &  97.86 &  97.70 & 4.84 & $+48.7/-22.0$ & $+13.9/-9.9$ \\
12 & O II    &   71.72 &  72.00 &  71.85 & 4.48  & &  \\
13 & He I    &   58.22 &  58.68 &  58.43 & 4.16  & &  \\
14 & H I     &   97.08 &  97.44 &  97.25 & 3.84  & &\\
15 & H I     &  102.42 & 102.70 & 102.57 & 3.84  & &\\
\hline
\end{tabular}\\
$^*$ Column 3 to 5 give the wavelength range and peak wavelength of each spectral line, Column 6 lists the
corresponding formation temperature, and the last two columns give the deviations of the variabilities and
gradients of final selected spectral lines (see Sec.\ref{sec_data} for more details).
\end{center}
\end{table*}

\end{document}